\documentclass[lettersize,journal]{IEEEtran}
\usepackage{amsmath,amsfonts}
\usepackage{algorithmic}
\usepackage{xcolor}
\usepackage{array}
\usepackage[caption=false,font=normalsize,labelfont=sf,textfont=sf]{subfig}
\usepackage{textcomp}
\usepackage{stfloats}
\usepackage{url}
\usepackage{verbatim}
\usepackage{graphicx}
\usepackage{cite}
\usepackage[authoryear]{natbib}
\usepackage{hyperref}
\usepackage{siunitx}
\def\BibTeX{{\rm B\kern-.05em{\sc i\kern-.025em b}\kern-.08em
    T\kern-.1667em\lower.7ex\hbox{E}\kern-.125emX}}
\usepackage{balance}

\begin{document}
\title{The Eye as a Window to Systemic Health: A Survey of Retinal Imaging from Classical Techniques to Oculomics}
\author{Inamullah, Imran Razzak, Shoaib Jameel}
\makeatletter
\def\@maketitle{%
  \newpage
  \null
  \vskip 1em%
  \begin{center}%
    {\LARGE \@title \par}%
    \vskip 1.5em%
    {\large
     Inamullah Inamullah\textsuperscript{a,*},
     Imran Razzak\textsuperscript{b},
     Shoaib Jameel\textsuperscript{a}\par}%
     \vskip 1em%
    {\small
    \textsuperscript{a} School of Electronics \& Computer Science, University of Southampton, SO17 1BJ, Southampthon, United Kingdom\par
    \textsuperscript{b}Mohamed bin Zayed University of
Artificial Intelligence
Abu Dhabi, United Arab Emirates \par
    \textsuperscript{*}Corresponding author: i1n23@soton.ac.uk \quad
    }%
  \end{center}%
  \par
  \vskip 1.5em
}
\makeatother

%\thanks{Manuscript created October, 2020; This work was developed by the IEEE Publication Technology Department. This work is distributed under the \LaTeX \ Project Public License (LPPL) ( http://www.latex-project.org/ ) version 1.3. A copy of the LPPL, version 1.3, is included in the base \LaTeX \ documentation of all distributions of \LaTeX \ released 2003/12/01 or later. The opinions expressed here are entirely that of the author. No warranty is expressed or implied. User assumes all risk.}

%\markboth{Journal of \LaTeX\ Class Files,~Vol.~18, No.~9, September~2020}%
%{How to Use the IEEEtran \LaTeX \ Templates}

\maketitle

\begin{abstract}

The unique vascularized anatomy of the human eye, encased in the retina, provides an opportunity to act as a window for human health. The retinal structure assists in assessing the early detection, monitoring of disease progression and intervention for both ocular and non-ocular diseases. The advancement in imaging technology leveraging Artificial Intelligence has seized this opportunity to bridge the gap between the eye and human health. This approach facilitates the unveiling of systemic health insights from the ocular system and the identification of non-invasive surrogate markers for timely intervention and detection. The new frontiers of oculomics in ophthalmology cover both ocular and systemic diseases, and are getting more attention to explore them. In this survey paper, we explore the evolution of retinal imaging techniques, the dire need for the integration of AI-driven analysis, and the shift of retinal imaging from classical techniques to oculomics. We also discuss some hurdles that may be faced in the progression of oculomics, highlighting the research gaps and future directions. 
\end{abstract}

\begin{IEEEkeywords}
Retinal Imaging, Retinal Vasculature, Oculomics, Systemic Diseases, Artificial Intelligence, Deep Learning, Retinal Biomarkers, Phenotypes, Genotypes
\end{IEEEkeywords}

\section{Introduction}

The retina, the crucial light-sensitive tissue at the back of the eye, is a treasure trove of insights for ophthalmologists, clinicians, and technicians. It plays a pivotal role in the early prediction and intervention of both ocular and non-ocular disorders. Retinal imaging, a key interpretation, encompasses a variety of techniques designed to visualise and analyse the retina \cite{ref1}. Colour Fundus photography (CFP) captures detailed, subtle changes in the ocular system, while optical coherence tomography (OCT) provides a cross-sectional view, and OCT angiography visualises blood flow. These modalities, serving as vital diagnostic tools, enable healthcare professionals to examine the intricate structure of the retina, including the macula, optic disc, and retinal vasculature. This non-invasive assessment allows for the early detection and monitoring of ocular diseases like diabetic retinopathy, glaucoma, macular degeneration, and other related diseases that can lead to significant vision impairments if left unattended \cite{ref2}. 
 
Beyond ocular health, retinal imaging also provides insight into detecting systemic diseases. The retina's unique accessibility and the visibility of its microvasculature make it a valuable indicator of overall health. For example, changes observed in the blood vessels and layers of the retina can trigger cardiovascular \cite{ref50},    neurodegenerative \cite{ref157}, and cerebrovascular diseases  \cite{ref2a}. That is why the eye serves as a window of overall health in the association of ocular and systemic health \cite{ref138}. High-resolution imaging can further reveal fine-grained features and act as a facilitator for timely intervention and personalized care. This contributes to improved patient outcomes and a better understanding of the interrelation between bodily systems \cite{ref3}.

Oculomics studies ocular biomarkers, using advanced imaging technologies to probe the eye in relation to other organ systems \cite{ref4} through shared physiological functions, embryological origins, and anatomical structures \cite{ref138}. It aims to establish correlations between these markers and systemic diseases \cite{ref3}. Growing evidence indicates that oculomics is significant in identifying clinical manifestations in ocular markers, such as retinal vessel patterns and neural layer thickness, which can reveal the presence of systemic diseases such as diabetes, cardiovascular diseases (CVDs), neurological issues, and stress-related disorders \cite{ref5}. This field is driven by three significant advances, including the widespread availability of high-resolution ophthalmic imaging, the emergence of big data, and the development of automated artificial intelligence (AI) software. Together, it has the potential to revolutionise disease management by providing a comprehensive and accessible approach to monitoring and maintaining human health \cite{ref6}.

AI is at its peak in the era of big data and parallel processing. The AI model is based on recorded data, and disciplines with data can easily participate in it. In the current epoch, it is hard to conclude that there may be fields in academia and industry that can thrive without data. Therefore, no discipline will be empty-handed when discussing data. AI models are now prevalent and deployed in every aspect of life \cite{ref7}. Both intrinsic and extrinsic deployments have positively impacted our society, especially in socioeconomic and socio-health aspects. Renowned computer scientist Dr. Knuth revealed that biology has at least 500 years of exciting problems to explore \cite{ref8}. This perspective encourages the computing and biomedical research communities to address challenges in life science collaboratively.

AI is transforming retinal imaging and ocular studies by automating and enhancing the analysis of complex visual data. AI algorithms, especially deep learning (DL), excel in identifying subtle patterns and abnormalities within retinal images that human observers might overlook \cite{ref9}. This automation significantly speeds up the screening process for ocular and non-ocular diseases, allowing for earlier detection and intervention. In addition, AI can enable the development of predictive models by evaluating the vast data set, encompassing the assessment of the individual patient's risk factors and anticipating disease progression, paving the way for personalised medicine in ophthalmology 
\cite{ref10,ref11}.

Eye diagnostic research is shifting from its classical origin to the current AI software tools. Now, it's beyond the classical methods, such as preprocessing, model training, and limited clinically relevant output. These limited circumstances hinder the quantification of retinal structures at the minute and micro levels to address the objective measurements, which are crucial for the efficacy of effective treatments and the monitoring of disease progression in the growing field of oculomics. Likewise, for meaningful research, high-quality and large health datasets comprising rich annotated data are crucial, which advances the role of statistical, computational, and AI-driven techniques \cite{ref12,ref13}. Furthermore, this framework can be further nourished by integrating retinal imaging with clinical information, namely electronic health records and genetic data, including diverse demographic and ethnic data.  However, translating this vision into practice remains a considerable challenge. Existing datasets often lack the necessary integration across epidemiological, clinical, and imaging domains, limiting the development of a unified infrastructure supporting oculomics research and clinical application \cite{ref4,ref14,ref15}.

Despite the increasing availability of large-scale datasets, significant limitations impede their practical use in oculomics. Even though there are several Retinal imaging repositories, including those of Moorfields Eye Hospital \cite{ref16}, UK Biobank \cite{ref17}, and the 10k Cohort \cite{ref18}, access to these resources is often restricted to trusted research environments due to anonymisation protocols. Moreover, much of the publicly available data lacks critical demographic and ethnic metadata, with estimates suggesting that approximately 74\% of these datasets do not include such information. As a result, even healthcare providers with extensive retinal image collections, ranging from ten thousand to ten million scans, face challenges in generating truly representative \cite{ref19}, integrative datasets. 

More broadly, a substantial gap exists in integrating retinal imaging data with molecular, clinical, and diverse datasets. Such integration is crucial for capturing the multi-dimensional nature of complex diseases \cite{ref20}. However, epidemiological models, such as cohort, longitudinal and population-based studies, have successfully mapped disease trajectories in other domains \cite{ref21}; their application to oculomics remains onerous \cite{ref4}. The combined burden of global communicable and non-communicable diseases has further highlighted the need for such inclusive, multi-modal datasets that account for diversity across ethnicity, age, and geography \cite{ref22}.

These shortcomings highlight the urgent need for a next-generation framework to unify fragmented data sources and enable deeper clinical insights. This includes robust epidemiological frameworks, large-scale imaging biobanks, and AI-enabled platforms harmonising disparate data types. Unified and multidimensional data frameworks, including molecular signatures, imaging biomarkers, and longitudinal clinical profiles, have the potential to provide a more holistic understanding of disease onset and progression. These frameworks, already successful in fields such as oncology and cardiology \cite{ref23}, are timely and necessary for their adoption in oculomics \cite{ref5}.

As a result of these converging needs and opportunities, oculomics has emerged as a leading frontier in multimodal healthcare research, with biomarkers to infer both ocular and systemic health \cite{ref24}. The retina is uniquely positioned for this purpose due to its rich vascular and neural architecture, which reflects broader physiological states throughout the body. As a transparent and non-invasive window into the vascular and neurobiological environment, retinal imaging allows early detection of systemic conditions. This makes the retina an ideal candidate for multimodal investigation in both research and clinical settings \cite{ref25}.

Recent developments in automated image analysis tools have further enhanced retinal imaging capabilities. The AutoMorph software developed by Zhou et al. \cite{ref1} enables a precise, large-scale extraction of morphological features from the retinal vasculature in a fully automated pipeline. This advancement significantly supports high-throughput oculomics research by offering standardised and scalable morphological analysis of retinal structures.  

In recent years, oculomics has emerged as a highly interdisciplinary domain, yet there remains a lack of cohesive literature synthesising its technological, clinical, and methodological advancements. Most existing reviews focus either on specific imaging techniques or isolated systemic diseases, without bridging the broader implications of AI-powered retinal analysis for integrated healthcare \cite{ref26,ref27,ref6,ref28,ref29}. Given the rapid development of retinal image processing tools, dataset accessibility, and AI-driven diagnostic models, a comprehensive review is now both necessary and timely. This survey aims to fill that gap by contextualising classical foundations, highlighting cutting-edge innovations like AutoMorph, and proposing a unified framework for future research. This survey paper presents a structured synthesis of advancements in retinal imaging, AI integration, and multi-modal health data. It highlights key breakthroughs, identifies current gaps in clinical translation, and offers a roadmap for future interdisciplinary collaborations in oculomics. Through this, we aim to support the development of robust, explainable, and scalable tools capable of revolutionising healthcare delivery across disciplines.

\section{Ocular Anatomy and Physiology \& Historical Perspective on Eye Disease Research}
\subsection{Ocular Anatomy and Physiology}

\begin{figure}[ht]
    \centering
    \includegraphics[width=\linewidth]{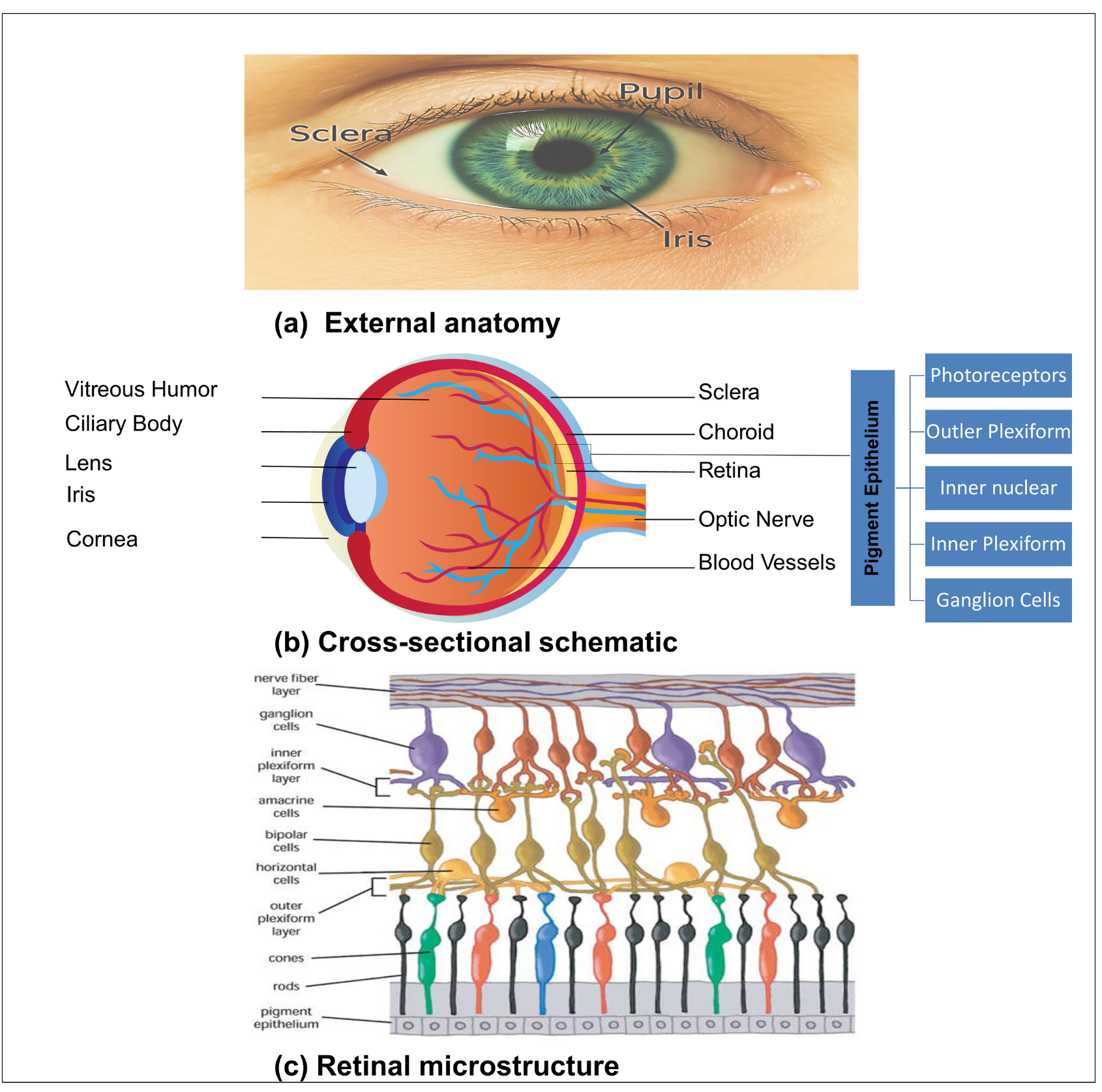} \caption{ Structural organization of the human eye at multiple anatomical levels.
(a) External anatomy of the eye showing the pupil (central opening), iris (colored ring), and sclera (white outer coat that maintains ocular shape and protection).
(b) Cross-sectional schematic illustrating the major internal structures including the cornea, iris, lens, vitreous humor, sclera, choroid, retina, blood vessels, and optic nerve. This orientation highlights the key optical and supportive components of the eye.
(c) Retinal microstructure depicting the layered neuronal architecture, including photoreceptors (rods and cones), bipolar, horizontal, and amacrine cells, ganglion cells, and the pigment epithelium.}
    \label{fig:1}
\end{figure}

The eye is a living optical system, and its role as a window into systemic health is grounded in its unique anatomy and physiology, where refractive, neuronal, and vascular components interact to enable vision and reflect disease processes. Each element of this system, illustrated in Figure \ref{fig:1}, is not only vital for sight, but is also sensitive to disturbance, which means that localised abnormalities often signal both ocular pathology and broader systemic disease.

%At the outer surface, the cornea, sclera, iris, and pupil regulate the entry of light and protect the delicate internal structures. As portrayed in Figure \ref{fig:1}a, the external anatomy of the eye includes the sclera, iris, and pupil, which together regulate light entry and maintain ocular integrity. The cornea, a transparent, curved dome, provides nearly three-quarters of the eye’s total refractive power. 

At the outer surface, the sclera, iris, and pupil regulate the entry of light and protect the delicate internal structures, as portrayed in Figure \ref{fig:1}a. The cornea, while anatomically part of the external coat of the eye, is transparent and therefore not readily visible in a superficial view. In cross-sectional orientation (Figure \ref{fig:1}b), the cornea is seen as a curved dome that provides nearly three-quarters of the eye’s total refractive power. Its highly ordered collagen fibrils maintain curvature and transparency \cite{ref32}, but when disrupted, as in keratoconus or corneal scarring, irregular astigmatism and impaired vision result \cite{ref29a}. Surrounding it, the sclera forms the opaque, fibrous shell that preserves the globe’s shape and stability. Excessive thinning and stretching of the sclera underlie high myopia, now a growing global public health concern \cite{ref29b}. The iris adjusts the pupil aperture to balance light entry: it widens in low-light environments and constricts in bright conditions, safeguarding the retina from excess light exposure.

Behind the iris lies the crystalline lens, a biconvex structure suspended by zonular fibers. The lens fine-tunes focus through accommodation, altering its curvature to shift between near and distant vision. With aging, the lens gradually hardens and its proteins aggregate, producing cataract, the world’s leading cause of reversible blindness \cite{ref29c}. Light then traverses the vitreous humor, a transparent gel that fills the posterior chamber. This medium stabilizes the ocular globe and transmits light without distortion, but it is not static. Over time, the vitreous liquefies and contracts, often detaching from the retina. Such posterior vitreous detachment can predispose to retinal tears and even retinal detachment, both of which pose a serious risk to sight.

The retina lies at the back of the eye and functions as the sensory core of vision. It is a multilayered neural tissue where photons are converted into electrical signals. As shown in Figure \ref{fig:1}c, the retinal microstructure consists of photoreceptors (rods and cones), bipolar, horizontal, and amacrine cells, ganglion cells, and the retinal pigment epithelium.Specialised photoreceptors, rods for dim-light and peripheral vision, and cones for colour perception and fine detail, initiate this process. Their output is integrated by horizontal, bipolar, and amacrine cells, which refine contrast and spatial information. Finally, ganglion cells collect these signals, and their axons converge to form the optic nerve, transmitting information to the brain’s visual cortex. Within the retina, the macula supports high-acuity central vision, while the peripheral retina detects movement and supports night vision \cite{ref30,ref31}. Cross-sectional analyses, such as those provided by OCT, reveal this layered architecture in vivo and demonstrate how dysfunction at any level, whether photoreceptor loss in age-related macular degeneration (AMD) or microvascular leakage in DR, directly alters visual capacity. Figure \ref{fig:1}b highlights these internal structures in cross-section, including the cornea, lens, vitreous humor, retina, choroid, and optic nerve

The retina is sustained by an intricate network of supporting tissues and circulations. The retinal pigment epithelium (RPE) nourishes photoreceptors, regulates waste clearance, and recycles visual pigments. The choroid, a dense vascular layer, supplies approximately 65\% of retinal oxygen and nutrients, while the retinal vasculature provides the remaining 35\% \cite{ref32}. This dual circulation reflects the enormous metabolic demand for the retina. Because retinal vessels are optically accessible, their geometry, summarised by traits such as vessel caliber, fractal dimension, tortuosity, and vessel density, provides quantifiable endophenotypes. These vascular signatures have been linked not only to local conditions such as DR but also to systemic disease, including hypertension, cardiovascular disease, stroke, and even renal dysfunction \cite{ref2a,ref32a}

The optic nerve forms the final bridge between the eye and the brain. Its vulnerability is twofold: localised ocular pathologies, such as glaucoma, that progressively damage retinal ganglion cell axons, while systemic neurodegeneration also manifests here. The thinning of the retinal nerve fiber layer (RNFL) has been observed in both Alzheimer’s and Parkinson’s disease \cite{ref32b}, suggesting that the eye can act as a biomarker of central nervous system health. This role, both in transmitting visual signals and in reflecting neurological status, cements its place in the investigation of systemic disease.

Collectively, the eye is not only an organ of sight, but also a mirror of health. Each component, from the cornea that bends light to the retina that transduces it into neural signals to the optic nerve that delivers these signals to the brain, offers unique diagnostic insight. Because these structures are accessible through modern imaging modalities, their disruption reveals both ocular disease and systemic pathology. This foundational understanding of anatomy and physiology has historically guided eye research and now underpins the transition to advanced imaging and oculomics, which will be explored in the following section on the historical development of ophthalmology.

\subsection{Historical Perspective of Eye Diseases Research: From Ancient Techniques to Modern AI-Driven Oculomics}

The study of the eye and associated diseases dates back to ancient civilisations, where early contributions focused on anatomy, diseases, and basic treatments. Over the centuries, the field of ophthalmology has evolved significantly, driven by the pressing need to understand, diagnose, and treat eye diseases. Today, AI integration has revolutionised the field, enabling unprecedented advances in diagnostics, treatment, and research. This section traces the historical journey of eye disease research, from ancient techniques to the era of classical AI and modern deep learning, culminating in the emerging field of oculomics.

\subsubsection{\textbf{Pioneer Surgical and observational Techniques}}

The study of the eye dates to ancient civilisations, where early contributions focused on ocular anatomy, diseases, and basic treatments. Cataracts were among the first documented eye diseases, dating back to around 600 BCE. The Sushruta Samhita, an ancient Indian medical text, described a surgical technique called ``couching'', where the opaque lens was inserted into the vitreous humour to restore some vision \cite{ref33}. However, insertion often led to complications, for instance infection and blindness, underscoring the limitations of early surgical techniques. 
Greek scholars Hippocrates (460–375 BCE) and Galen (129–216 CE) founded anatomical descriptions of the choroid, lens, and optic nerve. Their work provided the foundation for understanding ocular anatomy, but their reliance on animal dissection led to inaccuracies \cite{ref34}. Islamic scholars like Al-Zahrawi and Ibn al-Haytham improved surgical tools and advanced optical studies, connecting anatomy with light refraction. Ibn al-Haytham's Book of Optics emphasised the principles of vision and optics, enabling a framework for understanding refractive errors \cite{ref35}.

\subsubsection{\textbf{Early Breakthrough in the optical imaging}}

The development of the ophthalmoscope by Hermann von Helmholtz in 1851 was a transformative milestone. This device enabled clinicians to visualise the retina and optic nerve directly, facilitating diagnosing of conditions such as retinal detachment and optic neuritis \cite{ref36}. Although initial iterations of the ophthalmoscope required considerable skill to be effective, this limited its wider adoption in clinical practice.  The early twentieth century saw significant progress in tackling prevalent eye diseases such as trachoma and cataracts. Trachoma, a bacterial infection caused by Chlamydia trachomatis, was a leading cause of blindness in low-income regions. Public health campaigns focusing on hygiene, antibiotic use, and surgical interventions significantly reduced its prevalence. Nevertheless, these efforts were limited to localised outbreaks and lacked global coordination \cite{ref37}. 

\subsubsection{\textbf{Technological and optical imaging advancement during the Mid-20th Century}}

The invention of fundus photography in the 1920s enabled static documentation of the retina, providing a foundation for modern research of retinal disease. Although this innovation allowed for detailed visualisation, early cameras suffered from low resolution and were not widely accessible. Cataract extraction methods improved with the introduction of intracapsular cataract extraction. Due to the absence of intraocular lenses, visual outcomes were less than ideal, requiring patients to wear thick corrective lenses after surgery.

The mid-20th century saw rapid advances in imaging and treatment technologies. In the 1960s, fluorescein angiography allowed for dynamic retinal and choroidal vasculature imaging. It became a critical tool for diagnosing retinal vein occlusion and diabetic macular edema. However, the invasive nature of dye injection posed risks to patients with allergies or comorbidities \cite{ref38}. Laser treatments became the standard for managing proliferative DR (PDR) and retinal detachment. While effective, these treatments were limited to advanced stages of the disease and could not prevent disease onset. Tonometry was standardised as a reliable method for measuring intraocular pressure, enabling earlier detection of glaucoma. Tonometry alone cannot detect optic nerve damage, highlighting the need for additional diagnostic tools. \cite{ref39, ref40}.

\subsubsection{\textbf{The Emergence of Digital Era and AI in Ophthalmology}}

The digital era introduced technologies that significantly enhanced the precision and accessibility of ophthalmic diagnostics. Optical coherence tomography, invented in 1991, provided cross-sectional imaging of retinal layers with micrometre resolution . This technology revolutionized the management of macular degeneration, DR, and glaucoma, but early devices were expensive and required significant expertise \cite{ref41}.  Early computational models automated tasks like vessel segmentation and lesion detection. While these methods were groundbreaking, they were limited by small datasets and reliance on handcrafted features. Digital fundus photography enabled large-scale diabetic retinopathy screening programs. These initiatives demonstrated the feasibility of integrating technology into public health, but faced challenges in reaching rural and underserved populations \cite{ref32,ref11}. 
The integration of AI into eye disease research began with classical machine learning algorithms, which laid the foundation for automated analysis of retinal images and clinical data. Early models, such as Support Vector Machines (SVM), Random Forests, and k-Nearest Neighbors, were used for tasks like lesion detection, disease classification, and retinal vessel segmentation. These models relied on hand-crafted features extracted from CFP and OCT images, which limited their generalizability and scalability \cite{ref42,ref43,ref44,ref45}
For example, SVMs were employed to classify DR stages based on features, namely microaneurysms (MAs), haemorrhages (HMs), and exudates (EXs). These models faced challenges due to the varying image quality and complex retinal structures. Despite these limitations, classical Machine Learning (ML) algorithms demonstrated the potential of AI in automating ophthalmic diagnostics and reducing the burden on clinicians.

\subsubsection{\textbf{Early Deep learning models in Eyes research}}
The advent of deep learning, particularly Convolutional Neural Networks (CNNs), marked a paradigm shift in the analysis of ophthalmic imaging. Unlike classical ML, CNNs automatically learn hierarchical features from raw images, eliminating the need for manual feature engineering. This capability made CNNs highly effective for tasks like DR detection, glaucoma diagnosis, and age-related macular degeneration classification.  One of the earliest breakthroughs in deep learning for ophthalmology was the development of algorithms for the screening of diabetic retinopathy using fundus images. Studies demonstrated that CNNs could achieve a good diagnostic accuracy comparable to that of human experts \cite{ref46,ref47}. For example,  InceptionV3 and ResNet architectures were widely adopted for DR classification, achieving high sensitivity and specificity in detecting early-stage disease.  Similarly, CNNs were applied to OCT scans for the automated detection of retinal pathologies such as macular edema, choroidal neovascularization, and retinal detachment. These models improved diagnostic accuracy and reduced the time required for image analysis, enabling faster decision-making in clinical settings. 

\subsubsection{\textbf{The Rise of pre-trained AI Models and Multimodal Data Integration }}

CNNs revolutionized ophthalmic imaging, but their limitations in capturing long-range dependencies and spatial relationships led to the development of Vision Transformers(ViTs) \cite{ref48}. Inspired by the success of transformers in natural language processing, ViTs introduced self-attention mechanisms to image analysis, enabling better modelling of global context and spatial relationships within medical and optical imaging.  ViTs have been applied to tasks like DR grading, age-related macular degeneration classification, and retinal vessel segmentation, outperforming traditional CNNs in many cases. For example, Swin Transformers and Multiple Instance Learning ViTs (MIL-ViTs) have been used to detect subtle retinal changes associated with systemic diseases such as diabetes and hypertension \cite{ref49}. 

In addition, integrating multimodal data, which combines imaging, text, and molecular data, or may of the same modalities, has opened new frontiers in oculomics. For instance, AI models can now analyse retinal images alongside electronic health records and genomic data to predict disease progression and treatment outcomes. This multimodal approach has been particularly useful in studying complex conditions such as diabetic retinopathy and AMD, where genetic factors play an important role in disease pathogenesis \cite{ref50,ref51}. 

The application of AI extends beyond imaging to the realm of genomics and gene editing. Advances in next-generation sequencing and single-cell RNA sequencing have generated vast amounts of genomic and transcriptomic data, which can be analysed using AI to identify disease-associated genes and pathways.  For example, AI-driven genomic analysis has identified key genetic variants associated with AMD, DR and glaucoma \cite{ref52,ref53,ref54}. These insights have paved the way for gene editing technologies such as CRISPR-Cas9, which hold promise for treating inherited retinal diseases. AI models are also used to optimise gene editing protocols, predict off-target effects, and design personalised therapeutic interventions for individual patients \cite{ref55}. 

\subsubsection{\textbf{LLMS and Future AI in the OCULOMICS}}

The emergence of Large Language Models (LLMs) such as GPT-4 has further expanded the scope of AI in healthcare. LLMs can analyse vast amounts of textual data, including scientific literature, clinical notes, and patient records, to extract insights and generate hypotheses \cite{ref56}. Although in the field of oculomics, LLM is in the early stages and requires improvement, some results of the study show \cite{ref57,ref58} that the knowledge gained by LLM is generally taken by humans.  Although LLMS represent a cutting-edge branch of AI with growing potential in oculomics, it is equally important to understand the broader trajectory of AI as it intersects with medical and retinal research. The evolution of Artificial Intelligence, from its early origins to present-day applications, forms the foundation for modern techniques that now reshape retinal imaging. 

\subsubsection{\textbf{Overview on the Evolution of AI}}

\begin{figure*}[htbp]
    \centering
    \includegraphics[width=\linewidth]{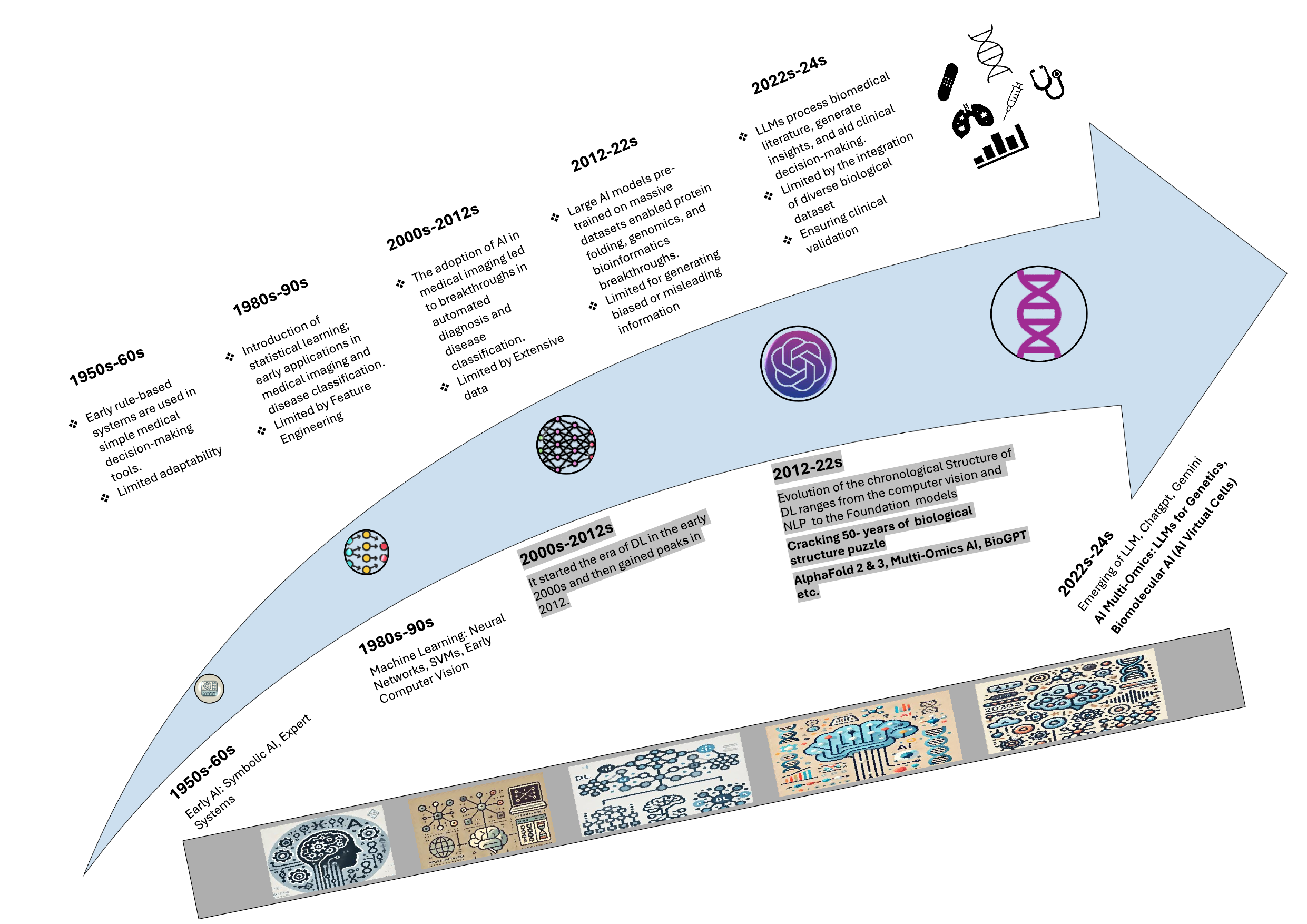}
    \caption{illustrates a timeline that outlines the development of artificial intelligence, starting with symbolic rule-based systems in the 1950s and progressing to modern deep learning, foundation models, and large language models. This evolution emphasises important technological changes that have impacted the overall AI field, providing a basis for today’s uses in healthcare and biomedical research.}
    \label{fig:4}
\end{figure*}

In a nutshell, the emergence of AI has reshaped the dynamics of almost every field, specifically healthcare. The short description, limitations, evolution, and relevance to medicine are shown in Figure \ref{fig:4}, and some related stuff is discussed here. In 1956, the emergence of AI took its main start from the Dartmouth College conference, which mainly consisted of symbolic AI, in other words, a rule-based system. However, due to the computational limit, the transition of the rule-based system into the classical ML occurred from 1966 to 1970. The wave of backpropagation was another milestone in 1980; despite the emergence of this technique, AI was not getting full attention due to the funding constraints, along with the expected results \cite{ref59}. The era 1980-1990 is considered a winter era for AI, and again, the domain is shifted to statistical and classical ML. The Chronology of LENET Architecture, which spans from 1990 to 2012, laid the foundation for the emergence of the DL model and paved a new horizon for the future of AI \cite{ref60,ref61}. The DL model is a seismic change that is now non-stop in the current era, and day-to-day, new algorithms and models are emerging. The transition of AI from a theoretical to a practical tool has changed the dynamics of the research, academia and industrial world. 

The capabilities of AI now extend from the basic models to the complex, whether supervised or unsupervised, transforming research, clinical practice, and industry alike. Within ophthalmology, this evolution intersects directly with the trajectory of retinal imaging. While classical imaging techniques laid the foundation for retinal diagnostics, they were largely limited to descriptive or handcrafted analyses. These approaches successfully characterised ocular lesions and vascular features but lacked scalability, sensitivity, and integration with systemic health. The rise of artificial intelligence has transformed this landscape, enabling automated segmentation, feature extraction, and predictive modelling at a scale unattainable with manual methods. Section \ref{Sec-3} explores how modern AI-driven methods build upon these classical foundations, translating retinal images into quantitative biomarkers with applications that extend beyond ocular disease to systemic health.

\section{Modern Techniques: The Rise of Oculomics}
\label{Sec-3}
As mentioned earlier, the eye disease is retained in the retina. The same is also connected and affects the important parts of the body, such as the brain, heart, kidneys, etc. This type of disease is called systemic disease, a condition that affects the entire body or several organs, not just one part. The linking of eyes with systemic diseases is coined oculomics \cite{ref1, ref4}. The potential behind oculomics has expanded further by leveraging the existing retinal imaging research with the advancement of multiple data modalities. This can be gained by revealing the medical modalities. The modalities used so far in the field of ophthalmology are OCT, fluorescence angiography (FA), and CFP to reveal structural and functional changes in the ocular system  \cite{ref2,ref3}. These modalities help provide enormous retinal imaging data and are widely used in the real world of medical vision. These imaging data emerged as the dataset used to mitigate the burden of eye diseases with respect to monitoring, therapeutic, prognostic, and diagnostic, respectively \cite{ref2}.  The datasets are used in different case studies individually and combined for monitoring and identifying a variety of eye diseases, such as DR, glaucoma, macular edema, and cataracts, etc.  The different retinal datasets are publicly available to detect and classify eye diseases on time. These datasets have been very helpful so far and are used for different objectives of classical ocular diseases. 

\begin{figure}[ht]
    \centering
         \includegraphics[width=\linewidth]{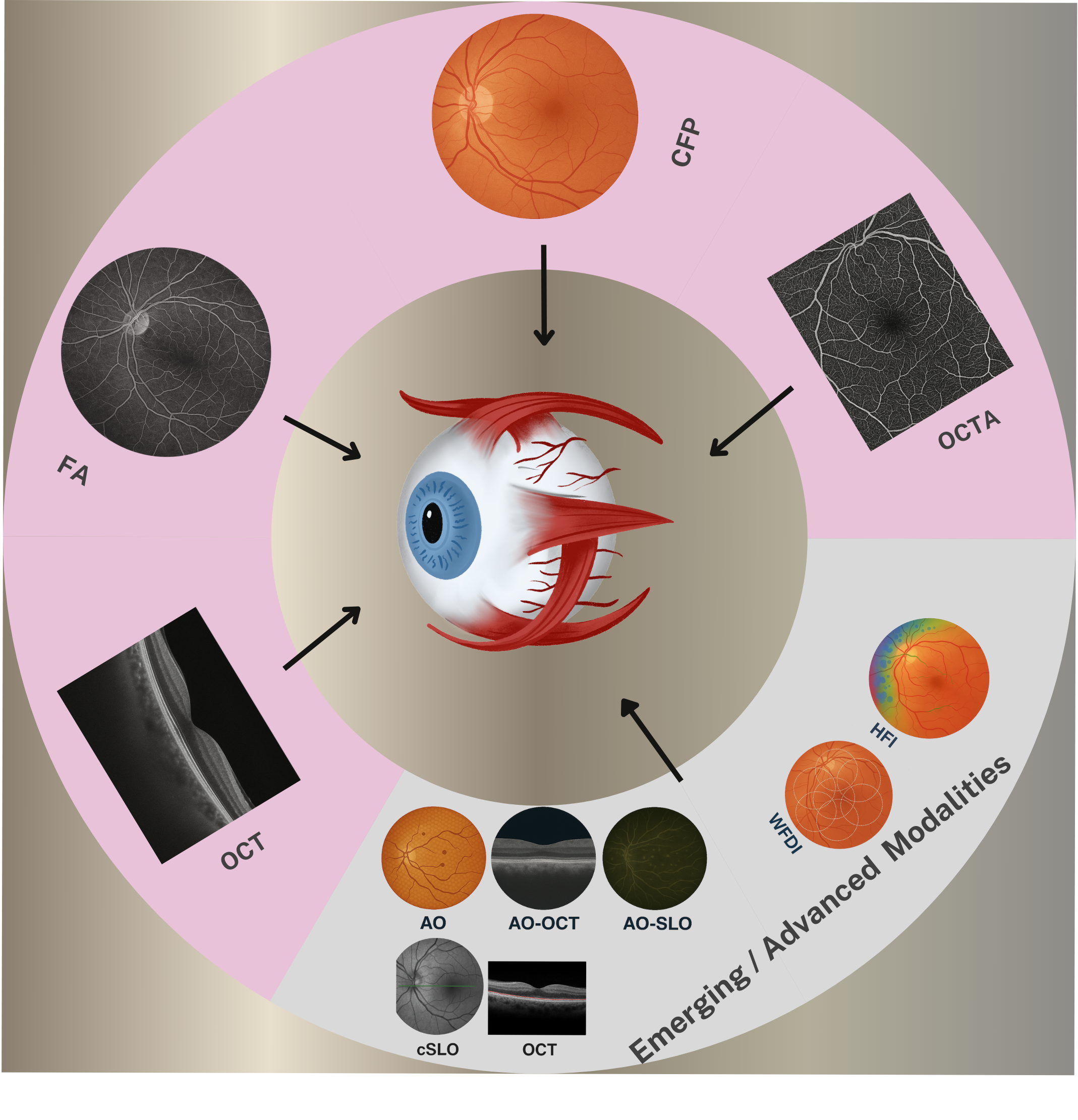}
    \caption{The schematic illustrates the spectrum of retinal imaging technologies. Modalities shown on a pink background (CFP, OCT, FA, OCTA) represent the core, widely adopted clinical tools that underpin current diagnostic practice. Modalities grouped on a white background (WFDI, HFI, AO, AO-OCT, AO-SLO, cSLO) reflect advanced or emerging approaches that extend the field of view, resolve cellular or biochemical detail, or enhance functional assessment. While these latter techniques remain largely research-focused, they point to the expanding potential of multimodal imaging for linking retinal structure with systemic disease processes.}
    \label{fig:2}
\end{figure}

\subsection{Retinal Imaging Techniques}

Retinal imaging technologies have evolved rapidly, trans-
forming our ability to capture both structure and function of
the eye at increasing resolution. Among the classical modalities, colour fundus photography remains one of the most widely applied tools in both clinical practice and large-scale epidemiological studies. Its strengths lie in simplicity, low cost, and broad availability, making it indispensable for screening conditions such as diabetic retinopathy, hypertensive retinopathy, and retinal vein occlusion. For decades, CFP has served as the standard tool in population-based studies, enabling quantification of vascular parameters such as caliber, tortuosity, fractal dimension, width, and focal narrowing. The rationale is that the retina, as the only part of the central nervous system visible non-invasively, provides direct access to microvascular and neural health \cite{ref63}. Modern image analysis platforms now extend their utility: SIVA (Singapore I Vessel Assessment), IVAN (Integrative Vessel Analysis), and VAMPIRE (Vascular Assessment and Measurement Platform for Images of the Retina) offer semi-automated measurements of vascular geometry \cite{ref63a}, while AutoMorph provides a fully automated, end-to-end pipeline for vessel segmentation and trait extraction \cite{ref1}. The major limitation of CFP remains its two-dimensional nature, which lacks depth resolution and reduces sensitivity to subtle intraretinal changes.

Optical coherence tomography introduced a step change in retinal diagnostics by allowing cross-sectional, micrometer-resolution imaging of the retinal layers. The progression from time-domain to spectral-domain OCT has greatly improved speed, resolution, and reproducibility, supporting the quantitative evaluation of the retinal sublayers \cite{ref61a,ref61b}. OCT is now the gold standard for detecting macular edema, glaucoma-related thinning, and AMD, although its cost and sensitivity to motion artifacts remain challenges. The functional extension, OCT angiography (OCTA), non-invasively visualizes retinal and choroidal microvasculature by detecting red blood cell motion. OCTA generates depth-resolved vascular maps and vessel density metrics that are particularly valuable in ischemic conditions and neovascular disease \cite{ref62}. Yet, OCTA can miss slow or turbulent flow, is prone to segmentation errors, and typically covers a narrower field than wide-field angiography.

Fluorescein angiography has long been regarded as a core imaging tool for assessing retinal vasculature. By injecting sodium fluorescein dye, FA highlights vascular leakage, non-perfusion, and neovascularization, making it essential in diagnosing diabetic macular edema, retinal vein occlusion, and neovascular AMD \cite{ref61c}. Despite its diagnostic utility, FA is invasive, time-consuming, and associated with risks such as nausea and, rarely, anaphylaxis, prompting the development of non-invasive alternatives like OCTA.

Beyond these established modalities, a number of emerging technologies extend the diagnostic horizon. Wide-field digital imaging (WFDI) captures up to 200° \cite{ref63b} of the retina in a single shot, proving especially useful in identifying peripheral changes in diabetic retinopathy and for neonatal screening \cite{ref63c}. Hyperspectral imaging (HFI) provides biochemical and metabolic information by measuring reflectance at multiple wavelengths, revealing signals such as hemoglobin oxygenation and amyloid deposits associated with Alzheimer’s disease \cite{ref63d}. These tools remain largely research-focused but hold promise for systemic disease characterization.

Adaptive optics (AO) should be understood as an enabling technology rather than a standalone clinical modality. By correcting ocular aberrations, AO achieves near-cellular resolution, and when coupled with imaging platforms such as scanning laser ophthalmoscopy (AO-SLO) or OCT (AO-OCT), it allows direct visualization of photoreceptors, retinal pigment epithelial cells, and fine capillaries in vivo. AO has provided unprecedented insights into inherited retinal diseases, including Stargardt disease, retinitis pigmentosa \cite{ref63e}, and cone–rod dystrophy \cite{ref63f}, and also shows potential for functional imaging by enabling the tracking of erythrocyte movement and local blood flow dynamics at the capillary level \cite{ref63}. However, its technical complexity and limited commercial availability currently confine AO to research laboratories rather than routine clinics. In parallel, confocal scanning laser ophthalmoscopy (cSLO) has enhanced classical retinal imaging, supporting modalities such as FA and fundus autofluorescence, with recent refinements including multicolor and ultra-widefield capabilities \cite{ref63g}. AO and cSLO extend imaging into both cellular and functional domains, complementing OCT and CFP within multimodal frameworks.

Today, the landscape of retinal imaging ranges from classical modalities such as CFP, FA, and OCT, which remain central to clinical care, to newer innovations including OCTA, WFDI, HFI, AO-enabled systems, and cSLO-based platforms that expand the scope of investigation into vascular, biochemical, cellular, and functional domains. As illustrated in Figure \ref{fig:2}, this spectrum can be organised into core modalities that underpin current diagnostic practice and emerging approaches that extend capabilities for research and multimodal integration. Each approach contributes unique strengths but also carries limitations, underscoring why multimodal combinations are increasingly applied. These limitations may be further compounded in patients with unstable fixation or ocular motility disorders such as nystagmus, where motion artefacts can compromise image quality and reduce the reliability of quantitative outcomes. Nevertheless, the convergence of classical and emerging technologies has generated an unprecedented volume of high-resolution ocular data, enabling the creation of large-scale repositories. These resources now play a central role in validating AI-driven pipelines and advancing oculomics toward clinical translation, as explored in the next section on retinal datasets.

\subsection{ Publicly Available Retinal Datasets}
In the present day, the data is used as a tool for competing, from small tech companies to giants and from developing to developed countries. Data is valuable and works as a gold mine in the epoch of GEN-AI. In this context, the available data is getting more attention if found in large amounts, specifically in the field of the medical sciences. This data can emerge as a dataset which can be used further in the real world of health research and drive innovations and discoveries \cite{ref19}.  Furthermore, gold-standard imaging techniques provide the opportunity for the creation of a dataset used as a benchmark for the validation and testing of real clinical trials in situ and in vivo. The advancement of retinal imaging techniques has created numerous publicly available datasets, which have become indispensable resources for researchers and physicians. These datasets are widely used for lesion detection, disease classification, and vessel segmentation \cite{ref65}. For example, datasets such as \textbf{STARE} \cite{ref66},\textbf{DRIVE} \cite{ref67},\textbf{DIARETDB1} \cite{ref68} ,\textbf{CHASE DB1}, \cite{ref69} ,\textbf{HRF} \cite{ref70},
\textbf{ E-ophtha} \cite{ref71} , \textbf{MESSIDOR}, \textbf{MESSIDOR-2} \cite{ref72} ,\textbf{KAGGLE-EyePACS} \cite{ref73} ,\textbf{IDRiD} \cite{ref74}, \textbf{KAGGLE-APTOS} \cite{ref75} , and \textbf{DDR} \cite{ref76} have been instrumental in developing algorithms for detecting and classifying diabetic retinopathy, segmenting retinal vessels, and identifying other ocular pathologies.

Collectively, these publicly available datasets have established the foundation for advancements in image-based ophthalmic research, particularly in the automated detection and classification of retinal diseases. However, the diagnostic requirements and imaging targets vary significantly across different ocular conditions. To understand how these datasets contribute to disease-specific model development, it is important to contextualise their relevance across major retinal disorders.

In this survey, therefore, we transition from the dataset overview into a focused discussion of key localized diseases: beginning with glaucoma, where structural deterioration of the optic nerve is a primary concern; followed by age-related macular degeneration, which involves degeneration of the macula; and finally diabetic retinopathy, where vascular abnormalities and lesion characteristics dominate. Each disease presents unique imaging biomarkers and machine learning challenges, necessitating tailored approaches in algorithm design and data representation.

\begin{figure}[ht]
    \centering
         \includegraphics[width=\linewidth]{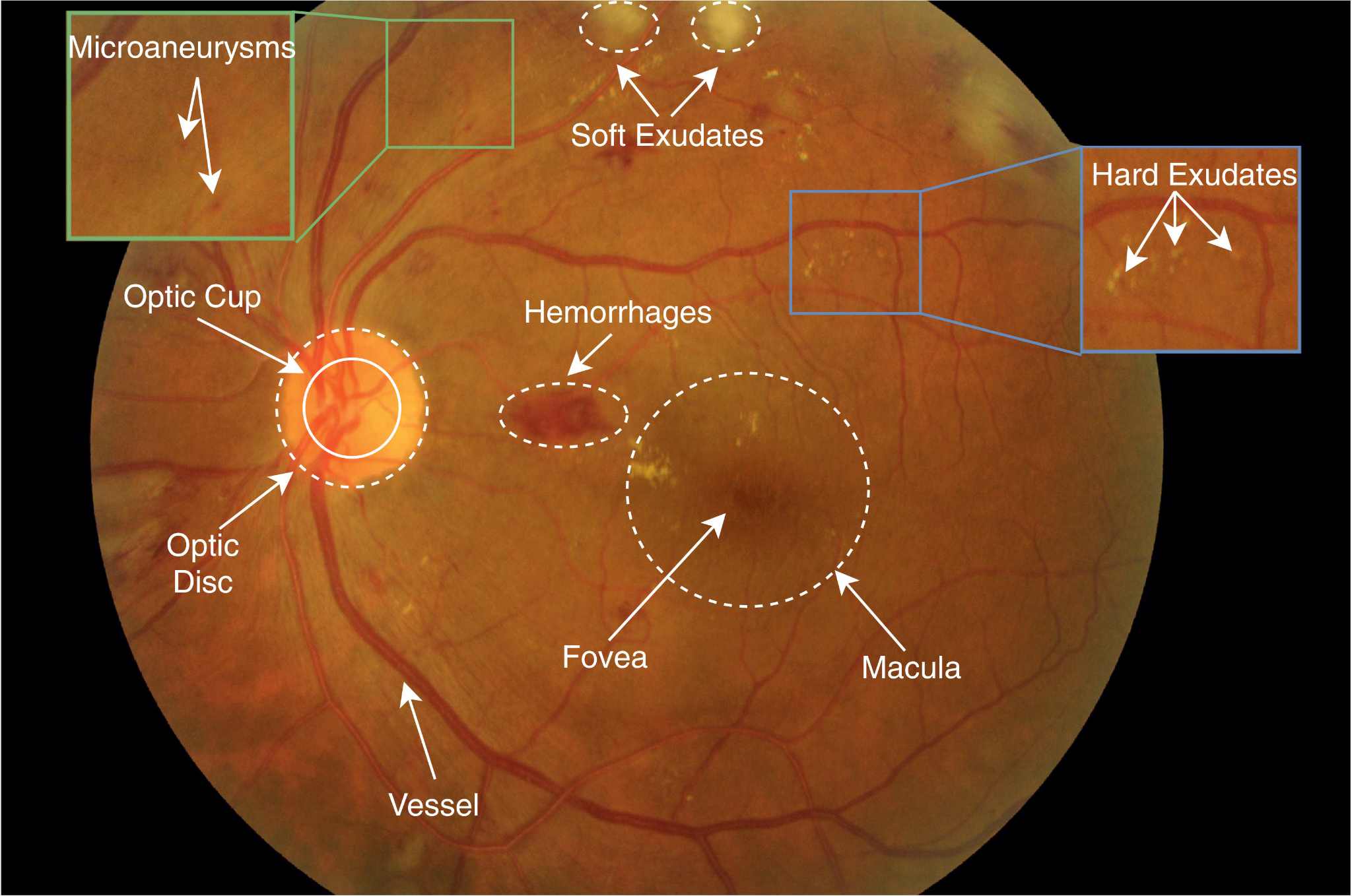}
    \caption{An annotated fundus image reproduced from \cite{ref65}, showing pathological markers including microaneurysm, haemorrhage, exudates, and anatomical regions like the fovea and optic disc.}
    \label{fig:3}
\end{figure}
\textbf{Glaucoma: } Glaucoma is a long-term and gradually worsening condition of the optic nerve that results in permanent loss of vision and stands as one of the primary causes of blindness worldwide. It affects an estimated 76 million people as of 2020 and is projected to impact over 111.8 million by 2040 \cite{ref77}. Elevated intraocular pressure is often linked to the disease, but many patients also experience normal-tension glaucoma, illustrating the condition’s complex origins. The disease affects both the structural and functional health of the eye, especially through the degeneration of retinal ganglion cells (GC) and harm to the optic nerve head.
Clinically, a key biomarker for diagnosing glaucoma is the cup-to-disc ratio, which quantifies the size of the optic cup relative to the optic disc and reflects the cupping of the optic nerve \cite{ref78,ref79}.

Two primary imaging modalities, CFP and OCT, are widely used to detect and monitor glaucomatous changes. OCT provides high-resolution cross-sectional views of the RNFL and GC complex, enabling the early identification of nerve fiber loss. In contrast, CFP offers a top-down view of the optic nerve head, allowing for disc cupping evaluation and broader topographic interpretation \cite{ref80}. \cite{ref81} demonstrated that the combination of OCT and CFP scans could improve accuracy in detecting structural glaucomatous damage by assessing the cup-to-disc ratio. In the last decade, deep learning has significantly reshaped the landscape of glaucoma diagnostics. Early ML models often relied on handcrafted features like texture and colour histograms to classify glaucoma cases \cite{ref82}. These models provided a baseline but lacked scalability and robustness in real-world scenarios. 

With the rise of CNNS, researchers shifted to end-to-end learning pipelines capable of automatically extracting hierarchical visual features\cite{ref10}. In a study,   \cite{ref83} developed a CNN-based glaucoma detection model trained on fundus images, reporting performance metrics on par with experienced ophthalmologists. The research conducted by \cite{ref84} established a multimodal deep learning framework that integrated Optical Coherence Tomography and visual field data. This approach effectively captured both the structural and functional aspects of the disease. Similarly, \cite{ref85} built on this research by employing deep regression models to forecast future visual field deterioration using only baseline OCT scans, providing a useful clinical resource for long-term monitoring.

More recently, attention-based architectures, specifically ViTs, have shown promise in retinal image analysis. Unlike CNNS, which rely on localised filters, transformer models can capture long-range dependencies and global context, making them suitable for capturing subtle glaucomatous patterns across the optic disc and RNFL regions.  \cite{ref86} applied transformer models to fundus images and reported superior classification performance and generalizability across datasets compared to conventional CNN-based baselines.

\textbf{Age-Related Macular Degeneration: } Age-Related Macular Degeneration is a name that suggests that it can affect the center vision part of the eye, the macula. Eyesight declines with the passage of time and can impact daily routine activities. It is complicated in the early and later stages, as the early stage shows minimal symptoms and can cause vision loss in the former stage \cite{ref87}. In the case of unattended patients, AMD cases will be projected to be 288 million patients by 2040 \cite{ref88}. Due to this, it poses an important public health challenge. Different case studies have used an AI-driven retinal diagnostic model to address this ailment. Mark and Andres designed the study \cite{ref89},\cite{ref90} to quantify and detect the abnormality of the macula in Fundus images using ML. This model helped to assess the risk of progression, along with the prediagnosis of AMD. 

A comprehensive study \cite{ref91} which leverages the ML algorithm, categorises the age-related macular degeneration stages along with the data-driven technique and develops a Computer-Aided Diagnosis framework. This model is classified into No, Intermediate, Dry, and Wet AMD, and the results are compared with those of the other studies. This model has a better result. Likewise, unlike typical machine learning, a deep learning algorithm is employed in different case studies of AMD patients. Due to their hierarchical, deep structure, the DL model gets an improved result by interacting directly with the hidden parameters and acting as an autonomous model. 

A deep learning-AI-driven solution has been developed for automated AMD detection and differential diagnosis, leveraging vision transformers, data augmentation, and Swin transformers to improve accuracy, efficiency, and cost-effectiveness \cite{ref92}. Likewise, a deep learning model integrates with the local outlier Factor algorithm's goal to classify the age-related macular degeneration from the OCT scans. The generalizations are tested on the unseen dataset, a Duke dataset, which performs very well in the early detection of AMD \cite{ref93}.

\textbf{Diabetic Retinopathy:} When insulin production is not enough in the human body, it can raise the glucose level in the blood and lead to diabetes. Type 1 and 2 both have a high socioeconomic effect. However, type 2 requires more attention as it leads to other chronic diseases associated with heart, kidney, nose, nerve and eye diseases \cite{ref94}.

DR is one of the complications directly related to the eye. The complication involves swelling of the retinal vasculature, causing blood and fluid to leak, and, at a later stage, it leads to vision loss if left unattended. The pathophysiology of the DR risk is distributed like this: the retinal vascular leakage causes fluid accumulation in the vitreous humour. Similarly, MA and HM are red lesions, and by the same soft and hard EX, a bright lesion disrupts normal vision, and the lesion advances in the shape of Neovascularization, leading to severe vision as portrayed in Figure \ref{fig:3} \cite{ref96}. The commonly occurring retinal lesions include MAs, intraretinal HMs, and venous beading (characterised by alternating areas of venous dilation and constriction). Additionally, intraretinal microvascular abnormalities, hard EXs (lipid deposits), and retinal neovascularization are also recognised as common types of lesions \cite{ref65}. Since these DR risks are distributed into binary and multiple classes, in multiple classes, there are often five classes, which are No, mild, moderate, severe and PDR \cite{ref95}. In contrast, in the binary, these classes are non-proliferative DR (NPDR), further subdivided into the mild, moderate, and Severe DR. Subsequently, the remaining stages are considered as PDR  \cite{ref97}. 

Over the past two decades, much research has been conducted to prevent DR from the classical algorithms to the modern ones \cite{ref98}. A tetragonal local octa pattern method was proposed to represent features of fundus images \cite{ref99}. Building on this, \cite{ref100}applied traditional ML algorithms such as SVM, decision trees, and random forests for classification. Likewise, the study conducted at \cite{ref101} employed the approach of Gabor wavelet encased with the Adaboost classifier to classify the DR. For the binary classification of a DR, a range of studies \cite{ref102,ref103,ref104,ref105} has been available using the CNN model, which has gained a significant result.

Furthermore, the binary and multi-classification of DR leveraging the DL model is evidenced in these studies. For instance, in \cite{ref106}, InceptionV3 was employed to detect DR using RGB and texture features.  \cite{ref107} developed binomial and multinomial classification models for fundus images using MobileNetV2, while \cite{ref108} utilised the DenseNet-121 model to identify DR from fundus images. Additionally, various architectures of EfficientNet were explored in \cite{ref109}. Some studies proposed hybrid models that combined DL models for feature extraction with traditional ML models for classification. For example, \cite{ref110}used InceptionV3 for representation learning in conjunction with an SVM for DR classification.

 Ensemble models (EMs) have long been recognised in medical image analysis as an effective way to improve classification accuracy and robustness by combining the outputs of multiple base learners \cite{ref9}. In the case of DR detection, \cite{ref112} demonstrated this approach with an ensemble of three CNNs, where AdaBoost was used to optimise the integrated predictions. More recently, attention has shifted toward transformer-based architectures, which employ self-attention mechanisms to capture long-range spatial relationships that CNNs often miss. The introduction of Vision Transformers \cite{ref48} marked a turning point, and subsequent studies have highlighted their value in medical imaging applications \cite{ref113}. Comparative evaluations confirm this advantage: \cite{ref114} found that Swin-Transformer and Vision-Transformer consistently outperformed CNN-based models such as EfficientNet and ResNet, as well as MLP-Mixer architectures. Building on this progress,  \cite{ref49} introduced the Multiple Instance Learning Vision Transformer (MIL-ViT), which was pretrained on large fundus datasets before fine-tuning for DR tasks. When tested in APTOS2019 \cite{ref75} and RFMiD2020 \cite{ref115}, MIL-ViT achieved superior accuracy compared to conventional CNNs.

An ensemble of transformer models was developed that integrates four distinct networks to evaluate the severity of DR \cite{ref116}. Furthermore, \cite{ref117} proposed a DR grading model that combines a vision transformer with residual attention. This model features two primary components: (1) a feature extraction block based on the transformer, which emphasises retinal haemorrhage and exudate areas, and (2) a grading prediction block employing residual attention to capture different spatial regions corresponding to various classes. For further insights into the application of transformers in medical imaging, readers are encouraged to explore recent studies \cite{ref118}, \cite{ref119}.

 These diseases and their detection are pivotal and are used in vivo and in vitro, and have established a foundation for the field of oculomics. However, diseases covered in the above data sets may not have the potential to provide deep insight into the underlying mechanism of genotype-wide association. These are primarily focused on phenotypic analysis, which limits their utility for genotypic studies. To fully realise the prospective of oculomics, the acquisition of datasets needs to be extended to include cohort-based studies that integrate clinical patient records, high-throughput data, and structural and functional imaging data. Such comprehensive datasets would enable researchers to explore the genetic and molecular underpinnings of ocular and systemic diseases, paving the way for personalised medicine. Importantly, one of the most promising directions for such integration lies in the automated quantification of retinal vascular morphology, as the retinal circulation directly reflects systemic vascular health and provides measurable traits that can bridge imaging, genetics, and multi-omics frameworks

\subsection{Automated Retinal Vascular Morphology: A Pathway to Oculomics}

The circulatory system plays a critical role in systemic health, and the dysfunction often manifests in the retina's microvascular structures. Retinal circulation, characterised by the unique accessibility and transparency, provides an invaluable opportunity to study vascular health non-invasively. Blood viscosity, vessel length, and vessel diameter are critical determinants of vascular resistance and have been closely examined within the retinal vasculature to uncover systemic disease mechanisms \cite{ref120}. In connection with these features and parameters, a vascular structure encased in the retinal circulation gained close attention to unveil the mechanism underlying these structures. In recent decades, interest has gone beyond the classical techniques of oculomics due to its relationship with systemic diseases, namely, both types of diabetes, cardiovascular diseases, and dementia \cite{ref121, ref122}. 

Automated analysis of retinal vascular morphology provides objective, quantitative measurements of the retinal circulation. Standardized protocols typically focus on the six largest arterioles and venules emerging from the optic disc, measured within predefined concentric regions, most commonly Zone B (0.5–1.0 disc diameters from the disc margin) and Zone C (1.0–2.0 disc diameters). From these measurements, the central retinal arteriolar equivalent (CRAE) and central retinal venular equivalent (CRVE) are derived, summarizing average vessel caliber, while their ratio (AVR) reflects the arteriovenous balance \cite{ref1}. Additional quantitative traits extend this characterization: fractal dimension describes the geometric complexity of the branching network \cite{ref122a}; tortuosity quantifies the curvature and deviation of vessels from a straight path \cite{ref122b}; vessel density captures the proportion of retinal area occupied by vasculature \cite{ref122c}; branching angle measures bifurcation geometry; and arteriovenous crossing patterns define vessel interactions \cite{ref32a}. These retinal vascular traits provide reproducible phenotypes that can be computed at scale, forming the foundation for downstream oculomics and multi-omics integration.

The microvasculature features of the eye in the in vivo assessment of imaging modalities allow a computerized approach to measuring the caliber of the retinal vessels of the arterioles, venules and the relationship between these. These measurements are the gold standard for oculomics, which can be correlated with systemic diseases\cite{ref50}. Studies showed that hypertension and atherosclerosis are associated with narrowing of the arteries\cite{ref120},\cite{ref123}, and DR is associated with dilation of the retinal veins\cite{ref124}. In addition, increasing retinal artery tortuosity is interrelated with hypercholesterolemia and hypertension\cite{ref125,ref126}.

Initially, retinal calibers were measured manually, a tedious process requiring human computation and traditional formulas. However, manual vessel segmentation often produces suboptimal results, which can hinder model performance. Over the past few decades, extensive research has been conducted to automate this process and extract retinal vascular features. Significant advancements have been achieved through various approaches, including feature-based methods, unsupervised graph-based techniques, and supervised deep learning models\cite{ref127,ref128,ref129,ref130,ref131,ref132,ref133}. These developments have led to the emergence of multiple software tools for clinical and research applications. However, most of these tools remain semi-automated, necessitating human intervention to correct vessel segmentation and identify features such as bifurcations, tortuosity, branching angles, artery/vein diameters, widths, and Fractal dimension.

To address these limitations, AutoMorph, a fully automated software package, was developed by \cite{ref1}, leveraging deep learning models to autonomously measure retinal calibers without human intervention. The AutoMorph pipeline consists of four key components: preprocessing, image quality grading, anatomical segmentation, and morphological feature measurement. The anatomical segmentation and morphological feature extraction stages play a crucial role in advancing oculomics, an emerging field that explores the relationship between retinal and systemic diseases. The third stage of AutoMorph enables precise segmentation of the optic disc/cup, vessels, arteries, and veins. The final stage extracts key vascular morphology metrics, including tortuosity (via distance measurement, squared curvature, and density) for each artery and vein, and fractal dimension to quantify vessel complexity. Additionally, it measures retinal vascular calibers for  CRAE and CRVE, AVR, and enables macular region-specific assessments.

With advancements in retinal imaging software, researchers can now efficiently quantify retinal vascular features, segment major arteries and veins, and capture the fine-grained structural details of smaller vessels with high precision. These developments have expanded the potential of retinal imaging beyond ophthalmic disease diagnosis, providing valuable insights into systemic health conditions \cite{ref5}. Increasing evidence suggests that retinal vascular morphology serves as a biomarker for various systemic diseases, including hypertension, diabetes, cardiovascular disease, and neurodegenerative disorders \cite{ref3}. By enabling automated high-throughput retinal vascular analysis, AutoMorph and similar technologies contribute to a deeper understanding of the intricate connections between retinal changes and systemic pathophysiology, reinforcing the growing role of oculomics in predictive medicine and disease monitoring.

\subsection{Evolution of Oculomics: From Classical to Future Frameworks}

The field of oculomics is transforming from traditional medical imaging to modern multi-modality and omics data integration. This shift is driven by three current paradigms: Automated AI software tools, big data, and high-resolution ophthalmic imaging. 

\begin{figure}[htbp]
    \centering
    \includegraphics[width=\columnwidth]{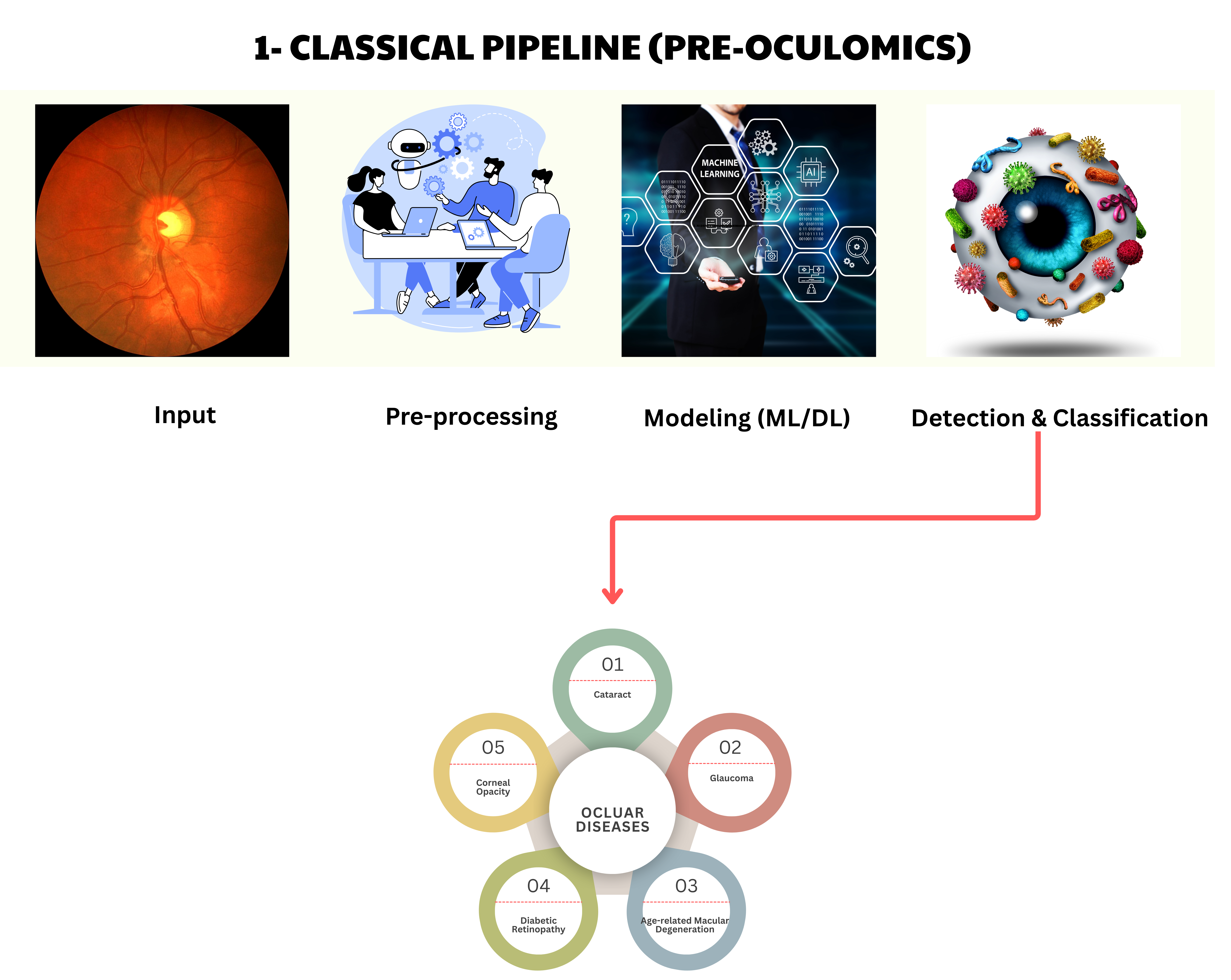}
    \caption{Before the rise of oculomics, computational pipelines for ophthalmology followed a straightforward path: Images were acquired as input for clarity, processed through pre-processing. The classical method went through feature engineering with a need for manual interpretation, including contrast enhancement, region of interest cropping and segmentation, and noise reduction. Then these were put in the model for training, testing and validating the model for the specific ocular diseases for machine learning and early deep learning modelling. This work is in the classification, detection and prediction of different diseases. These systems focused almost entirely on ocular diseases such as cataract, glaucoma, age-related macular degeneration, diabetic retinopathy, and corneal opacity. While they laid the foundation for computer-assisted diagnosis, they were limited in scope, largely handcrafted, and confined to the eye itself without considering wider systemic health.}
    \label{fig:5}
\end{figure}

Figure \ref{fig:5} represents the classical paradigm of eye research before oculomics emerged. In this era, diagnostic pipelines were built on conventional image preprocessing methods such as contrast enhancement, region-of-interest cropping, and manual noise reduction. Machine learning or early deep learning models were applied to handcrafted features, such as texture patterns, vessel widths, and colour intensity, extracted from fundus photography or optical coherence tomography. The outputs of these pipelines were typically limited to binary or multi-class classification (e.g., disease present or absent), with low interpretability, of localised optometry diseases, and minimal systemic relevance. This approach lacked scalability and failed to capture the full complexity of vascular and neural changes in the retina that could indicate broader systemic health conditions.

\begin{figure*}[htbp]
    \centering
    \includegraphics[width=\linewidth]{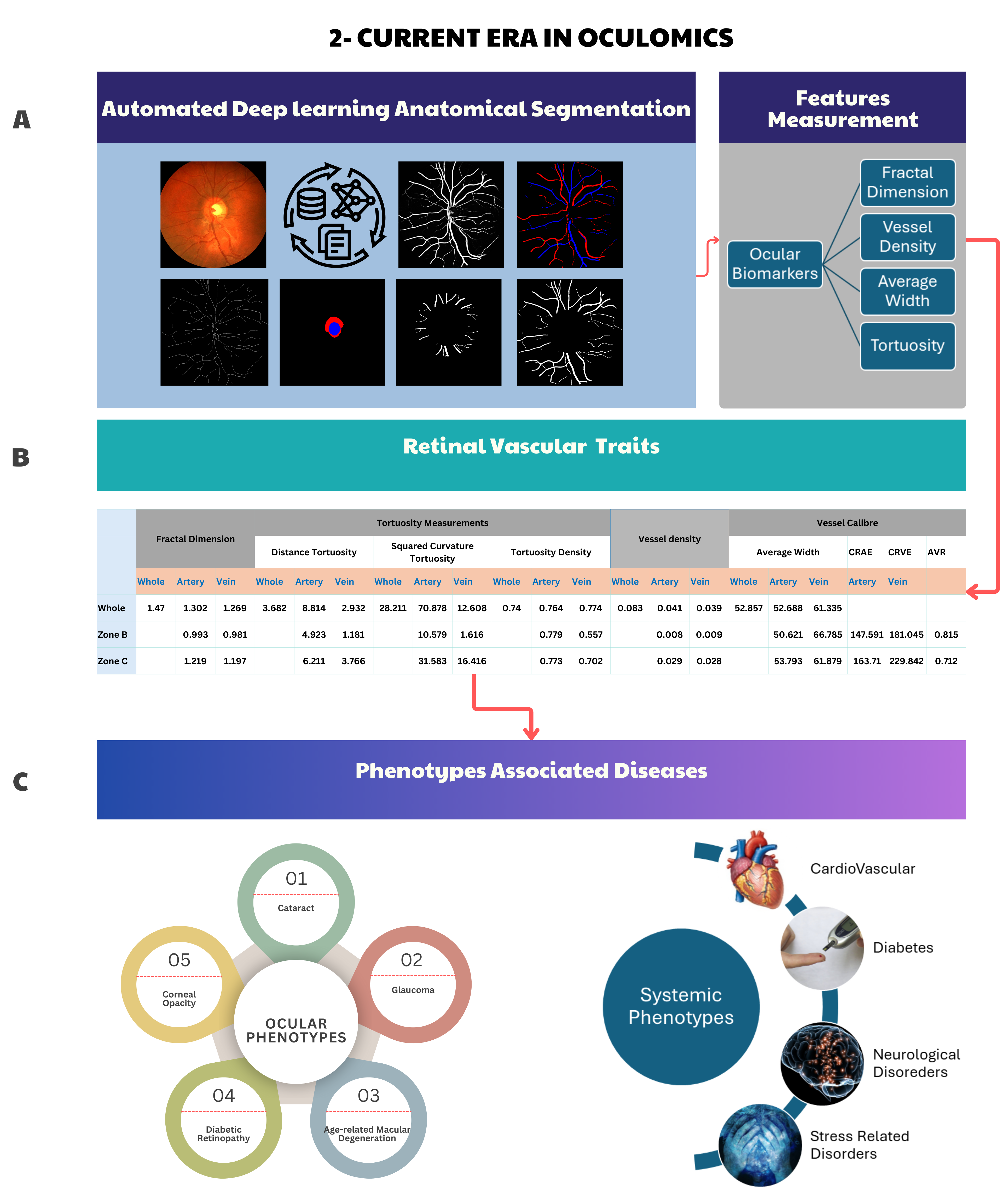}
    \caption{(A) Left-to-right workflow: Fundus images undergo automated quality assessment, followed by vessel segmentation (binary masking), artery/vein classification, skeletonization, and disc/cup delineation. Peripapillary measurement zones (B \& C) are defined as concentric annuli centred on the optic disc for standardized caliber and trait quantification. 
(B) Vascular Biomarkers: The system extracts quantitative vascular metrics, including dimensions, fractal dimension, vessel density, and tortuosity indices for both arteries and veins. Outputs, generated using AutoMorph on CHASE DB1 (image 007-L) \cite{ref69}, are reported in micrometres calibrated to the device's field of view. (C) Transition to Systemic Biomarkers: By standardizing and scaling retinal phenotypes, AI facilitates the association of ocular conditions (e.g., diabetic retinopathy, glaucoma, age-related macular degeneration) with systemic diseases (cardiovascular, metabolic, neurodegenerative), transitioning from handcrafted descriptors to traits suitable for population studies and enabling multi-omics integration.}
    \label{fig:6}
\end{figure*}

Figure \ref{fig:6} represents the current phase of oculomics, where deep learning automates the entire process of retinal vascular analysis. Starting with quality control, fundus images are either accepted or rejected before progressing through vessel probability mapping, binary segmentation, artery–vein separation, skeletonisation, and disc/cup segmentation. Standard epidemiological zones (B and C) are then overlaid around the optic disc to provide consistent reference frames for vascular measurement. From these outputs, quantitative biomarkers such as CRAE, CRVE, AVR, fractal dimension, vessel density, and multiple tortuosity indices are derived in a standardized way. Unlike the handcrafted approaches of earlier decades, these pipelines generate reproducible traits at scale, making them suitable for population cohorts and cross-study comparisons. By translating images into structured, quantitative endophenotypes, automated analysis has transformed the retina into a measurable biomarker source rather than a descriptive image. This stage is crucial in the evolution of oculomics. It bridges classical imaging, which remained focused on ocular disease, and future frameworks, where retinal traits are integrated with genetics, proteomics, metabolomics, and digital health data. In other words, Figure \ref{fig:6} does not just show technical automation; it marks the point where retinal imaging becomes scalable, reproducible, and ready for systemic integration.

Recent studies further illustrate how oculomics is expanding beyond descriptive pipelines into integrative, mechanistic frameworks. For instance, \cite{ref50,ref133a} demonstrated how high-resolution fundus datasets can be combined with clinical records and Mendelian randomization to show that changes in vascular morphology, such as arteriolar narrowing or reduced fractal dimension, are not merely correlates but causally implicated in major vascular events including stroke and myocardial infarction. This move from association to causation provides evidence that retinal traits can function as mechanistic biomarkers rather than statistical surrogates. Deep phenotyping cohorts, such as the UK Biobank \cite{ref17} and the 10K \cite{ref18} study, further strengthen this approach by embedding retinal imaging within multi-layered datasets that include lipidomics, metabolomics, immune profiling, body composition, and many more. Complementing these resources, \cite{ref133b} has extended deep phenotyping even further by integrating genetics, transcriptomics, metabolomics, immune and microbiome profiling, lifestyle data, continuous physiological monitoring, and imaging into a prospective cohort of more than 28,000 participants, alongside the development of foundation AI models trained for disease prediction. In another study, analysis from the 10K project has shown how deeply phenotyped cohorts can also reveal sex-specific dynamics of biological ageing, with machine learning models generating system-specific biological age scores that predict age-related disease risk beyond chronological age \cite{ref133c}. These studies reveal that retinal traits can be contextualised within broader biological signatures of health and ageing, providing a richer framework for interpreting systemic variation across populations. 

In parallel, plasma proteomic research has highlighted how organ-specific proteomic ages, particularly those of the brain and immune system, are strong predictors of Alzheimer’s disease and longevity, with hazard ratios exceeding 3.0 even after adjustment for conventional risk factors \cite{ref133d}. Such findings offer a mechanistic model for how retinal imaging could be linked with proteomic and genomic data to stratify risk for neurodegeneration and systemic aging. Disease-specific pipelines also illustrate the breadth of this vision: Eye2Gene, trained on multimodal retinal imaging, achieved top-five diagnostic accuracy of nearly 84\% across 63 inherited retinal disease genes, demonstrating how foundation-style phenotyping can accelerate genetic diagnosis \cite{ref133e}. Similarly, DeepSLE, a deep learning system to detect Systemic lupus erythematosus (SLE), has shown that autoimmune disease, specifically SLE, can be detected from fundus photographs using AI with performance comparable to serological screening, extending oculomics beyond cardiometabolic and neurodegenerative domains into immunology \cite{ref133f}. 

\begin{figure*}[htbp]
    \centering
    \includegraphics[width=\linewidth]{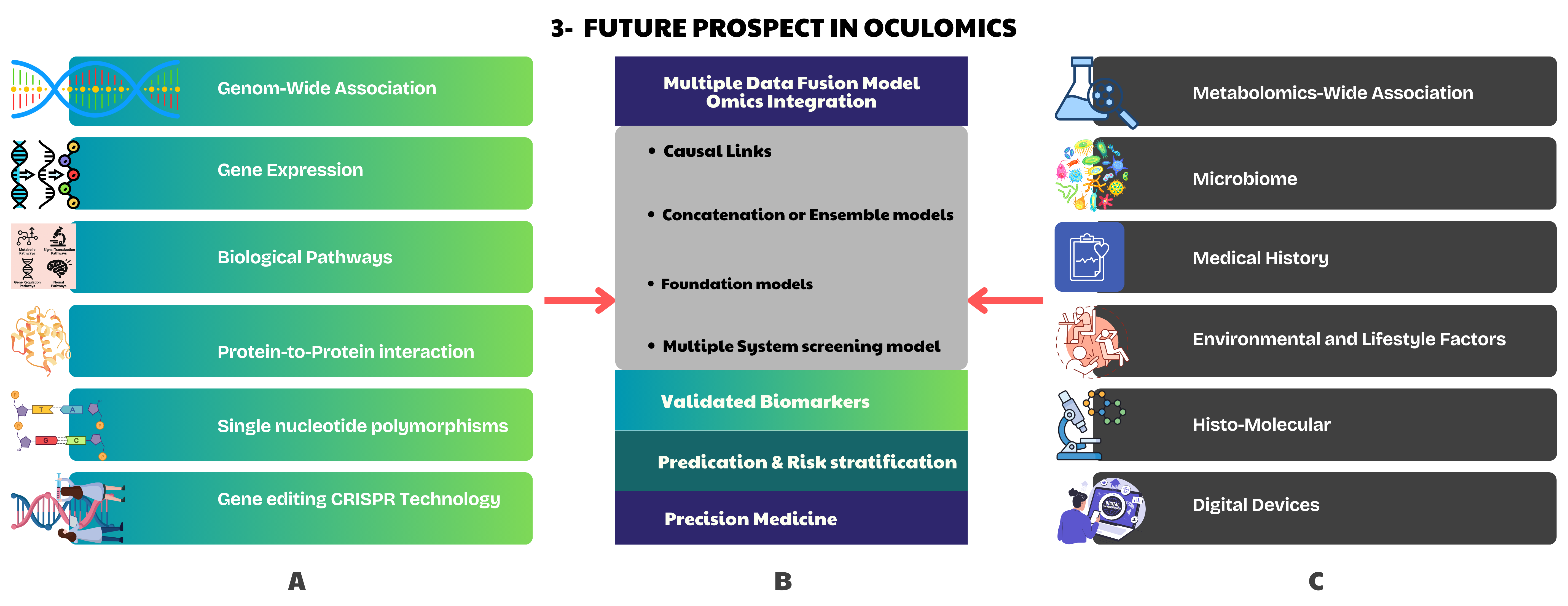}
    \caption{Future prospects in oculomics, toward validated biomarkers and precision medicine.
(A) Genomic \& molecular inputs: GWAS, gene expression, pathways, protein–protein interactions, single-nucleotide polymorphisms (SNPs), and gene-editing constraints provide a mechanistic context for retinal traits. (C) Clinical, environmental \& digital context: metabolomics, microbiome, medical history, lifestyle and environmental factors, histo-molecular readouts, and data from digital devices. (B) Computational core: multi-modal data fusion supports four complementary strategies, (i) causal inference (e.g., Mendelian randomization) to separate correlation from causation; (ii) multimodal ensembles/concatenation for robust prediction; (iii) foundation/next-generation phenotyping models (e.g., Eye2Gene for IRDs) to learn generalizable retinal representations; and (iv) multi-system screening models that jointly profile ocular and systemic phenotypes. Outputs include validated biomarkers, risk stratification, and decision support for precision medicine. Example systems span automated vascular phenotyping (AutoMorph), phenotypic screening with genetic validation for vascular events, deep learning for systemic autoimmunity (DeepSLE), and IRD gene prioritization (Eye2Gene).}
    \label{fig:7}
\end{figure*}

Together, these examples capture the trajectory summarised in Figure \ref{fig:7}. Current studies are beginning to show how causal inference can elevate retinal biomarkers from correlates to mechanisms, how multimodal concatenation of retinal and clinical data improves prediction, and how foundation-style models like Eye2Gene can compress complex phenotypes into actionable diagnostic outputs. At the same time, multi-system screening models such as DeepSLE preview a future where a single retinal image could support simultaneous risk profiling across cardiovascular, neurological, and immune conditions. Looking forward, the integration of retinal imaging with genomic-wide associations, plasma proteomics, metabolomics, digital devices, and longitudinal clinical data, processed through ensemble and foundation AI models, will define the next-generation framework of oculomics. In this vision, the inputs from genomics and clinical/digital streams (Figure \ref{fig:7}A and \ref{fig:7}C) converge in advanced computational architectures (Figure \ref{fig:7}B) to produce validated biomarkers, risk stratification tools, and decision-support systems that move oculomics firmly into the domain of precision medicine.

This figure underscores the ambition of oculomics to evolve from a diagnostic subfield of ophthalmology into a central node in precision medicine. Figures \ref{fig:5} through \ref{fig:7} collectively offer a visual synthesis of this paper’s thematic arc. They represent the chronological and conceptual development from basic retinal diagnostics, through current AI-driven phenotyping, and toward a highly integrative, multi-omics-enabled future. 

This transition also aligns with the paper's overall structure: beginning with classical foundations, moving through tools like AutoMorph, and culminating in the future of AI-integrated oculomics. By placing these stages adjacent to each other, the infographics demonstrate the evolution of the field from organ-centric diagnostics to a comprehensive approach, where the eye serves as a window to understanding human health at both the cellular and molecular levels. Such a framework supports the central argument of this paper: oculomics is not merely a subfield of ophthalmology, but a rapidly evolving interdisciplinary platform for next-generation healthcare innovation. Building on the current state of AI-enabled retinal imaging, the subsequent section expands into the systemic implications, demonstrating how retinal biomarkers are increasingly used to infer cardiovascular, neurological, and metabolic diseases.

\clearpage

\section{Linking Eye and Systemic Diseases}
\subsection{The Eye as a Mirror of Systemic Health}
The eyes are considered a window for the mirror of the soul of health. What does this mean to us? This means that looking at the eyes of the person can be interpreted, and the other person's concealed attitudes, thoughts, and emotions can be guessed \cite{ref134}. The same is true; the eyes are the only organ in the human body that acts as a window to see the world, making them unique organs to care for. In connection with other non-communicable diseases like rheumatic, neurodegenerative, hypertensive and diabetic Mellitus, eyes will help us to detect them in a non-invasive manner and act as biomarkers for these fatal diseases. 

The World Health Organisation estimated that 2.2 billion people in the world are suffering from eye diseases and are being prevented to the tune of one billion of them. Therefore, other than an ophthalmologist, researchers and scientists are keenly interested in unveiling the hidden patterns and mechanisms to analyse with the latest research, emulsified with the holistic approach to care for our eye health and save the whole body \cite{ref135}.  In addition, a recent study revealed that the eye also manifests a variety of stress diseases in which oxidative stress is crucial. This leads to the onset of different degenerative and chronic disorders such as cancer, atherosclerosis, coronary arteries, kidney, neurological, Respiratory, and Rheumatoid arthritis diseases  \cite{ref3,ref96}

 The advancement in retinal imaging technologies, such as fundus photography, OCT, and OCTA, has not only improved the diagnosis of sight-threatening ocular diseases but also opened the door to using ocular markers as indicators of systemic conditions. These biomarkers can anticipate disease onset and serve as non-invasive surrogates for risk stratification and treatment monitoring \cite{ref4}. In essence, the eye is both a mirror and a window to systemic health, reflecting numerous chronic disorders. Building on this perspective, the following section reviews systemic diseases where retinal changes are most evident, with a particular emphasis on age-related conditions. 
\subsection{Systemic Diseases Linked to Retinal Changes}

\subsubsection{ \textbf{Cardiovascular Diseases}}

Blood vessels captured in retinal images can be visualised noninvasively, and subtle differences in their caliber and branching patterns provide valuable insights into systemic vascular health and future cardiovascular risk \cite{ref5}. The ability to stratify risk is central to preventing CVD, which remains a leading cause of mortality worldwide. Current clinical scores, such as the Framingham Risk Score (FRS) and pooled cohort equations, rely on factors like age, smoking, body mass index (BMI), cholesterol, glucose, blood pressure, and sex, often supplemented by blood tests \cite{ref136}. However, these approaches face practical limitations. For instance, cholesterol values are not always available, with one study noting they were missing for nearly 30\% of patients in a 10-year risk calculation \cite{ref137}. Alternatives, such as using BMI in place of lipids, have been proposed, but they only partially capture metabolic risk. In this context, the retina offers an attractive non-invasive surrogate, providing a direct view of the microvasculature and its relationship to systemic disease \cite{ref15}.

Early population studies consistently reported that narrower arterioles and wider venules were associated with hypertension, coronary artery disease, and atherosclerosis \cite{ref138}. These cross-sectional observations highlighted the eye’s potential as a cardiovascular biomarker but did not establish causality. More definitive evidence came from large-scale longitudinal analyses. A pooled individual-participant meta-analysis of 20,798 people across six cohorts showed that wider venular caliber independently predicted incident stroke, with a hazard ratio of 1.15 (95\% CI: 1.05–1.25) per $20\,\mu\text{m}$ increase, whereas arteriolar caliber was not significantly associated (HR 1.00, 95\% CI: 0.92–1.08) \cite{ref138a}. Importantly, incorporating venular caliber into standard risk models reassigned more than 10\% of individuals at intermediate stroke risk, underscoring the added predictive value of retinal vascular traits. Similarly, hypertensive retinopathy,characterised by vessel narrowing, haemorrhages, and exudates, not only reflects chronic blood pressure elevation but also signals a higher likelihood of stroke, heart attack, and heart failure \cite{ref139}. These findings underscore that the vascular features of the retina can serve as validated predictors, not merely correlates, of cardiovascular outcomes.

More recently, advances in artificial intelligence have expanded the predictive scope of retinal imaging. A deep learning model was induced by \cite{ref139a} leveraging retinal vascular traits to predict major adverse cardiovascular events, including circulatory mortality, myocardial infarction, and stroke, with a C-statistic of 0.75 to 0.77, comparable to FRS. In a similar study, \cite{ref15} showed that fundus photographs alone could predict 5-year cardiovascular events with an AUC of 0.70, matching the performance of the composite SCORE calculator. In people with diabetes, the addition of retinal parameters to polygenic risk scores for coronary artery disease improved prognostic accuracy over a 10-year period \cite{ref139b}. Other biomarkers, such as the retinal age gap, a measure comparing biological and chronological age derived from retinal images, have also been shown to predict cardiovascular mortality in multiple cohorts \cite{ref139c,ref139d}. 

Integrative approaches further highlight the potential of oculomics. Combining fundus photographs with conventional clinical risk factors or dual-energy X-ray absorptiometry improved CVD risk prediction by 2–3\% \cite{ref139e,ref139f}. Optical coherence tomography angiography-based vascular parameters have also been investigated as markers of coronary artery disease \cite{ref139g}. Although their additional predictive value over established risk factors remains modest, particularly in patients with advanced diabetic or hypertensive retinopathy, they demonstrate the breadth of retinal imaging modalities that can contribute to cardiovascular risk assessment. 

Finally, large-scale AI studies integrating genetics have begun to provide causal insights \cite{ref50}. In the UK Biobank, deep learning models predicted 10-year risks of stroke, myocardial infarction, and chronic kidney disease with AUC values (area under the receiver operating characteristic curve) of 0.74–0.77, performing on par with or better than traditional clinical scores. Importantly, genome-wide association studies and Mendelian randomization analyses confirmed causal links between retinal vascular traits and cardiovascular outcomes, while also identifying reverse effects of chronic kidney disease on retinal microcirculation \cite{ref133a}. These advances mark a shift from descriptive correlations to validated, mechanistic pathways, placing the retina at the forefront of precision cardiovascular medicine.

\subsubsection{\textbf{ Diabetes and Diabetic Retinopathy}}

Diabetes, or in other words, Diabetes Mellitus, is one of the most pressing challenges in the socio-health scenario of the 21st century.  It has experienced a dramatic rise in the incidence rate for the past three decades.  The international diabetes Federation estimates that, to the tune of 536.5 million, the adult age group suffered from diabetes in the year 2021. This number could reach 738.2 million by 2045 \cite{ref142}. 

This disease association is characterised by hypercalcemia, which further exacerbates other complications like CVDS, kidney failure, and vision loss and imposes substantial burdens on the economy and healthcare system \cite{ref94}. Microvasculature complications damage the small blood vessels linked to the different organs and diseases like Diabetic Nephropathy, Neuropathy and Retinopathy \cite{ref143}. Microvasculature plays an important role in maintaining tumorigenesis and organ health, and the human eye, i.e., the retinal fundus, serves as a window for in vivo assessment \cite{ref50}. OCT and OCTA have facilitated the non-invasive assessment of the retinal microvasculature \cite{ref144}.

Studies have shown that individuals with diabetes exhibit alterations in retinal vascular caliber, including wider venular diameters, which are associated with an increased risk of developing diabetes and its complications. A comprehensive analysis with 18,771 participants indicated that a larger retinal venular diameter is linked to a higher risk of developing Type-2 diabetes over a median follow-up duration of 10 years, even when controlling for possible confounding factors \cite{ref145}.

Moreover, retinal texture analysis has emerged as a potential tool for the early detection of diabetes-related retinal changes. A recent study utilising OCT-derived texture characteristics found that changes in retinal texture are concurrent with biological retinal changes, suggesting that texture analysis could serve as a biomarker for the early diagnosis of DR \cite{ref142}. Circulating biomarkers have also been found, which also play a role in predicting diabetic complications. Higher levels of immunocomplexes and immunoglobulin M anti-cardiolipin antibodies have been observed in diabetes patients with vascular complications, suggesting their potential as biomarkers for DR \cite{ref146}.

Similarly, the progression of DR is closely linked to systemic metabolic control, evidenced by some recent studies, which underline the implication of advanced glycation end products (AGEs) and oxidative stress in the development of DR. The accumulation of AGEs in the retinal vasculature triggers the complication of inflammation and endothelial dysfunction; by the same oxidative stress, it worsens the cellular damage \cite{ref96, ref147}. These complications, on the one side, contribute to the progression of DR and, on the other side, reflect the systemic disease related to diabetes, like neuropathy and nephropathy.  For example, the presence of DR has been associated with a higher risk of diabetic kidney disease, underscoring the interconnectedness of microvascular complications \cite{ref148,ref149}.

\subsubsection{ \textbf{Neurodegenerative Diseases}}

The retina, an extension of the central nervous system, offers a unique opportunity to visualize brain health in vivo. Acting as a “window to the brain,” retinal imaging has revealed structural and functional alterations that mirror pathological processes in Alzheimer’s disease (AD), Parkinson’s disease (PD), multiple sclerosis (MS), and even cerebrovascular disease. This dual role, as both a mirror and a biomarker, has increasingly positioned oculomics as a powerful approach to detect early neurodegenerative changes before they manifest clinically \cite{ref3}. 

In Alzheimer’s disease, thinning of the retinal nerve fiber layer and the loss of the ganglion cell layer (GCL) are among the most consistent findings. Studies have shown that individuals at high genetic risk for AD exhibit reduced macular thickness even before cognitive symptoms emerge \cite{ref150}. Patients with AD also present with significantly thinner central macula compared with healthy controls \cite{ref151}. Retinal microvascular abnormalities, such as narrower venules and increased tortuosity, further suggest a vascular contribution to AD-related pathology \cite{ref152,ref153,ref154}. AI-based models analyzing retinal features have achieved promising results: one deep learning study using over 12,949 fundus images from 648 AD patients achieved diagnostic accuracies ranging from 0.80 to 0.92 \cite{ref154a}. However, while these biomarkers are strongly correlative, their independent predictive value beyond established markers such as amyloid PET or cerebrospinal fluid assays remains to be validated. 

In Parkinson’s disease, retinal changes also mirror underlying brain pathology. OCT studies have revealed retinal thinning and dopaminergic cell loss, both correlating with disease severity and progression \cite{ref155,ref156}. Beyond these structural alterations, functional markers are also being explored. Eye-tracking studies have identified prolonged saccadic latency and reduced response accuracy in PD patients, with machine learning models achieving AUC values of 0.73–0.93 in distinguishing patients from healthy controls \cite{ref156a,ref156b}. Although these findings remain largely correlative, they highlight the potential of combining structural retinal imaging with functional eye-movement analysis to improve early, non-invasive monitoring of neurodegenerative disease. In Alzheimer’s disease, deep learning applied to eye-tracking data has likewise shown that visual attention heatmaps can distinguish patients from controls with performance metrics of 0.84 accuracy and 0.90 AUC \cite{ref156c}. This provides early evidence that non-invasive eye-movement behaviours could serve as functional biomarkers for cognitive decline.

In multiple sclerosis, evidence is stronger and more validated:  Longitudinal OCT studies consistently demonstrate thinning of the retinal RNFL and  GCL as biomarkers of disease activity and disability progression \cite{ref157}. These structural markers have been further supported by machine learning approaches: support vector machines, ensemble classifiers, and recurrent neural networks applied to OCT parameters have achieved AUC values ranging from 0.82 to 0.88 for MS diagnosis and prediction of progression \cite{ref157a}. More recently, \cite{ref157b} demonstrated that CNNs trained in OCT thickness maps of sweep source, augmented with synthetic data using a deep generative adversarial network, could distinguish newly diagnosed MS patients from controls with near-perfect sensitivity and specificity (both 1.0). The retinal structures contributing most to discrimination included the ganglion cell layer plus inner plexiform layer (GCL+), the extended ganglion cell complex including RNFL and inner plexiform layer (GCL++), and the complete retina. These findings establish retinal imaging as both a structural and a computational biomarker platform to monitor MS, complementing its role in other neurodegenerative disorders.

Finally, cerebrovascular disease links further underscore the eye–brain connection. Longitudinal population-based studies consistently show that retinal vascular traits are predictive of stroke risk. In the Rotterdam Study, wider venular caliber was independently associated with both cerebral infarction and intracerebral hemorrhage, with hazard ratios of 1.28 (95\% CI: 1.13–1.46) and 1.53 (95\% CI: 1.09–2.15), respectively, while associations with narrower arteriolar caliber were weaker and borderline significant \cite{ref157c}. The Multi-Ethnic Study of Atherosclerosis (MESA) extended these findings to a diverse U.S. cohort, demonstrating that narrower arteriolar caliber (HR = 3.01, 95\% CI: 1.29–6.99) and retinopathy in non-diabetic individuals (HR = 3.07, 95\% CI: 1.17–8.09) were independently associated with incident stroke, even after adjustment for atherosclerotic markers such as carotid intima-media thickness and coronary calcium \cite{ref157d}(Klein et al., 2010). In Asian populations, the Singapore Malay Eye Study showed that larger venular caliber (HR = 3.28, 95\% CI: 1.30–8.26) and retinopathy signs predicted stroke events, and that the inclusion of retinal microvascular measures improved risk prediction and reclassification beyond established factors \cite{ref2a}. The decrease in retinal fractal dimension was similarly associated with stroke risk, with an odds ratio of 1.80 for the venular and 2.28 for the arteriolar networks \cite{ref157e}. More recently, large-scale AI studies have strengthened these associations. A UK Biobank study of 45,161 participants reported that each standard-deviation change in vascular traits, including vessel density, caliber, and branching complexity, was associated with a 9.8–19.0\% change in stroke risk. Incorporating these measures into conventional models improved predictive accuracy, increasing the area under the ROC from 0.739 to 0.752 \cite{ref157f}. Similarly, the DeepRETStroke framework, trained on hundreds of thousands of retinal photographs across multiple countries, predicted incident stroke with an AUC of 0.901 by detecting silent brain infarctions, outperforming conventional predictors across diverse populations \cite{ref157g}.

These validated findings illustrate how subtle retinal vascular changes can forecast cerebrovascular accidents years in advance, making them one of the most clinically actionable domains of oculomics. Collectively, these examples show how the retina serves as a biological bridge between the eye and brain. While many associations remain exploratory, validated pathways in multiple sclerosis and stroke demonstrate the clinical promise of oculomics. Ongoing advances in AI, multimodal imaging, and wearable eye-tracking technologies are expected to strengthen this bridge, enabling earlier detection, more accurate risk stratification, and deeper insights into the mechanisms of neurodegenerative disease.

\section{Future Directions in Oculomics}
Despite advancements in the field of oculomics from the classic to the modern. The emergence of AI models and their multi-faceted functions opens a door for interdisciplinary programs to dive into and solve socioeconomic and socio-health problems. Many of the studies included in this review paper tackle the localised and associated diseases of the eyes. However, this field still needs more attention to tackle these problems in one go. 

Here are future directions to heighten this field: design and develop a robust, monolithic, reliable and reproducible Automatic diagnosis system (ADS) to act as a life-saving tool for clinical and industrial research.

\begin{enumerate}
\item It is important to mention that the current retinal imaging datasets mainly focus on phenotypic characteristics, such as lesion segmentation, disease classification, and vascular analysis. However, they lack genetic and molecular data, limiting our understanding of ocular disease progression and systemic correlations. This gap is particularly evident in diseases such as DR, AMD, and glaucoma, where genotypic variations play a critical role in disease progression and treatment response. A major limitation of existing datasets is that they do not capture the progression of gene expression changes at different disease stages. For example, in DR, we currently lack information on:
\begin{enumerate}
 \item Gene expression variations associated with early, moderate, and proliferative DR stages.

 \item Patterns of arterial, venous, and capillary alterations and their correlation with gene activity.

 \item Molecular pathways linking DR to cardiovascular risks, including blood pressure, BMI, lipid levels, and systemic inflammation markers.
\end{enumerate}
This cannot be covered by the existing public dataset, such as those used in \cite{ref66,ref67,ref68,ref69,ref70,ref71,ref72,ref73,ref74,ref75,ref76}. Expanding epidemiological studies, such as cohort and longitudinal approaches, are essential and will cover the genetic sequencing data from the same patients, allowing for cross-comparison of gene expression, imaging biomarkers, and clinical factors. Tracking the same patients over time will help us to understand the underlying mechanism and result in a reliable outcome. 

Likewise, this will create retinal imaging datasets that include multi-omics data, such as genomics, transcriptomics, proteomics, and metabolomics, which are crucial to uncovering disease mechanisms at a systems level. This approach will allow researchers to study the genetic inclination to retinal diseases, track disease progression through molecular signatures, and establish links between ocular and systemic health. This chain will be interlinked with fostering interdisciplinary collaboration in areas like ophthalmology, data science, computer science, bioinformatics, molecular biology, and life sciences, opening a door for the success of oculomics. Similarly, the easily accessible multi-dataset (open access) will be a gold mine for institutional and healthcare research. The policymakers engage to ensure the practical implementation of data privacy and security to innovate the advancement in oculomics. 

    \item Expanding the retinal imaging capabilities to work with a diverse range of imaging, namely OCT, OCTA, AO-OCT, fundus and infrared imaging. The Automorph package still has the same limitation: they are only suitable for the fundus images and cannot be relied on if its modalities are other than this. A diverse range of images should be integrated, which will be suitable for all the modalities of retinal images. This improvement will help us with the subtle retinal changes in the cellular and vasculature network of the eyes and act as ADS for localised and systemic diseases like hypertension, diabetes and neurodegenerative conditions. 
    
    \item Foundation model and domain shifting are the future of medical image analysis; these two are scarce in oculomics. Domain shifting is hard to correlate with the CFP as it is well suitable for the OCT, and this can be achieved by shifting histoimaging of different tissues into the domain of eye tissue. However, the question arises as to why this model cannot be applied to all the modalities of optic systems. However, there is promise in this methodology, and it could be a popular approach in the future. Subsequently, the foundation model is trained on the enormous, diverse dataset and, by the same fine-tuning through a large number of tasks, produces reliable models. This model is found rare in the study of the retina and can be quite helpful in promoting ophthalmology. For example, the retina is linked with different diseases, including CVD. Finding the biomarkers for CVD with respect to retinal imaging. Creating a different model outcome like the features of carotid arteries correlates with the retinal imaging, the same for the serum lipidomics, sleep monitoring and oral microbiome dataset. Exploring the dataset directly connected with CVD and is associated with retinal imaging through the statistical model. Once a linkage is found and at the end, we add these models via the foundation model. This mechanism will be another way of contributing to eye research. In addition, this will maximise the transition of the research setting to routine clinical usage and could be expandable to portable retinal imaging devices to work as an ADS tool, which will also be helpful for the easily accessible remote regions of third-world countries. 
    
    \item The black box experience is the Elephant in the room, which is still challenging; however, the explainable AI solves some parts of it and goes further. This field itself needs attention. However, they have solved real-life puzzles. In connection with this, Explainable AI can also heighten the medical vision domain and help us identify which morphological patterns are responsible for systemic diseases through the DL system. 
    
    \item LLM is another buzzword in today's era and is trained on almost the world's puzzle. However, there is still room for oculomics, and we should focus on applications like the NYUTron \cite{ref159}, which is trained on 750,000 patients for routine clinical notes. Expanding to eye research can mitigate the burden on health practitioners. Likewise, the LLM medical specialist model Llava-Med \cite{ref160} and PathChat \cite{ref161} demonstrate promising biomedical and histopathology imaging capabilities. There is a dire need for oculomics research communities to design an application specific to retinal systems to uncover complex patterns in one go.

\item Finally, it pertains to mention that despite the inclination of AI in the medical domain and the advancement in AI-based medical and ocular applications, they are still in the translational stage, and their benefits in clinical practice remain challenging. To fully utilise the potential of AI in oculomics medical trials, these key challenges need to be addressed.
\begin{enumerate}
\item The integration of multiple data modalities as discussed earlier

\item The development of transparent and explainable AI models encased with the code of conduct for data sharing and model validation.

\item The collaboration between the AI researcher, clinician and healthcare practitioner ensures the care of AI-based tools relevant to the needs of stakeholders of the public health sector. 

\item The oculomics domain expert should be familiar with the rapid advancement of AI technology and tools. This will enable him to navigate the world and solve real-time problems in a routine clinical environment. 

%\textcolor{blue}{Despite rapid progress, several technical and clinical challenges remain. Patient-specific factors such as unstable fixation in nystagmus can introduce motion artefacts in OCT and OCTA, limiting the reliability of quantitative biomarkers. Similarly, while adaptive optics has enabled cellular-level imaging, its potential for quantifying blood flow dynamics is still under development. Imaging outcomes in conditions such as retinal vein occlusion also require further validation, as current evidence remains limited. Beyond structural imaging, functional modalities (e.g., hyperspectral or fluorescence lifetime imaging) are emerging but not yet widely adopted. Finally, future work must also clarify how imaging-derived biomarkers can guide novel interventions such as genome editing in inherited retinal diseases. Addressing these challenges will be critical to translating oculomics into validated, clinically actionable tools.}

\end{enumerate}
\end{enumerate}
\section{Conclusion}

Our detailed survey presents a broad exploration of the evolution of retinal imaging, tracing its path from classical techniques such as fundus photography and early machine learning approaches to the emerging era of oculomics powered by artificial intelligence. The transition from traditional vision-based diagnostics to deep phenotyping frameworks demonstrates the expanding role of the retina as a non-invasive biomarker for systemic health. Critical gaps remain despite significant advancements, such as the development of automated tools, the rise of multimodal AI frameworks, and the emergence of vision transformers. The disparity of datasets, the integrational paradigm lack between clinical, genetic, and imaging data, and limited demographic diversity constrain real-world deployment. Additionally, AI systems' interpretability and adaptation to heterogeneous clinical environments remain open challenges. As oculomics advances, it offers a unique opportunity to unify molecular, phenotypic, and clinical insights. This field could further nurture early diagnosis, risk prediction, and personalised healthcare by enriching an inclusive, longitudinal, and ethically governed research infrastructure. Sooner or later, the eye may serve not only as a diagnostic tool but as a window into systemic health and human biology at large.

\bibliographystyle{agsm}
\bibliography{PCC_Ref}

@article{ref1,
  title={AutoMorph: automated retinal vascular morphology quantification via a deep learning pipeline},
  author={Zhou, Yukun and Wagner, Siegfried K and Chia, Mark A and Zhao, An and Xu, Moucheng and Struyven, Robbert and Alexander, Daniel C and Keane, Pearse A and others},
  journal={Translational vision science \& technology},
  volume={11},
  number={7},
  pages={12--12},
  year={2022},
  publisher={The Association for Research in Vision and Ophthalmology}
}

@article{ref2,
  title={Advances in multimodal imaging in ophthalmology},
  author={Ringel, Morgan J and Tang, Eric M and Tao, Yuankai K},
  journal={Therapeutic Advances in Ophthalmology},
  volume={13},
  pages={25158414211002400},
  year={2021},
  publisher={SAGE Publications Sage UK: London, England}
}

@article{ref2a,
  title={Retinal microvascular changes and risk of stroke: the Singapore Malay Eye Study},
  author={Cheung, Carol Yim-lui and Tay, Wan Ting and Ikram, M Kamran and Ong, Yi Ting and De Silva, Deidre A and Chow, Khuan Yew and Wong, Tien Yin},
  journal={Stroke},
  volume={44},
  number={9},
  pages={2402--2408},
  year={2013},
  publisher={Lippincott Williams \& Wilkins Hagerstown, MD}
}

@misc{ref3,
  title={Oculomics--The eyes talk a great deal},
  author={Honavar, Santosh G},
  journal={Indian Journal of Ophthalmology},
  volume={70},
  number={3},
  pages={713},
  year={2022},
  publisher={Medknow}
}

@article{ref4,
  title={Insights into systemic disease through retinal imaging-based oculomics},
  author={Wagner, Siegfried K and Fu, Dun Jack and Faes, Livia and Liu, Xiaoxuan and Huemer, Josef and Khalid, Hagar and Ferraz, Daniel and Korot, Edward and Kelly, Christopher and Balaskas, Konstantinos and others},
  journal={Translational vision science \& technology},
  volume={9},
  number={2},
  pages={6--6},
  year={2020},
  publisher={The Association for Research in Vision and Ophthalmology}
}

@article{ref5,
  title={Retinal vessel diameters and function in cardiovascular risk and disease},
  author={Hanssen, Henner and Streese, Lukas and Vilser, Walthard},
  journal={Progress in retinal and eye research},
  volume={91},
  pages={101095},
  year={2022},
  publisher={Elsevier}
}

@article{ref6,
  title={Oculomics: Current Concepts and Evidence},
  author={Zhu, Zhuoting and Wang, Yueye and Qi, Ziyi and Hu, Wenyi and Zhang, Xiayin and Wagner, Siegfried K and Wang, Yujie and Ran, An Ran and Ong, Joshua and Waisberg, Ethan and others},
  journal={Progress in Retinal and Eye Research},
  pages={101350},
  year={2025},
  publisher={Elsevier}
}

@article{ref7,
  title={Computational medicine: past, present and future},
  author={Lyu, Lan-qing and Cui, Hong-yan and Shao, Ming-yi and Fu, Yu and Zhao, Rui-xia and Chen, Qiu-ping},
  journal={Chinese journal of integrative medicine},
  pages={1--10},
  year={2021},
  publisher={Springer}
}

@article{ref8,
  title={A review on machine learning aided multi-omics data integration techniques for healthcare},
  author={Bansal, Hina and Luthra, Hiya and Raghuram, Shree R},
  journal={Data Analytics and Computational Intelligence: Novel Models, Algorithms and Applications},
  pages={211--239},
  year={2023},
  publisher={Springer}
}

@article{ref9,
  title={Deciphering the impact of diversity in CNN-based ensembles on overcoming data imbalance and scarcity in medical datasets: A case study on diabetic retinopathy},
  author={Hassan, Saima and Belhaouari, Samir Brahim and Amin, Ibrar and others},
  journal={Informatics in Medicine Unlocked},
  volume={49},
  pages={101557},
  year={2024},
  publisher={Elsevier}
}

@article{ref10,
  title={Deep learning in estimating prevalence and systemic risk factors for diabetic retinopathy: a multi-ethnic study},
  author={Ting, Daniel SW and Cheung, Carol Y and Nguyen, Quang and Sabanayagam, Charumathi and Lim, Gilbert and Lim, Zhan Wei and Tan, Gavin SW and Soh, Yu Qiang and Schmetterer, Leopold and Wang, Ya Xing and others},
  journal={Npj Digital Medicine},
  volume={2},
  number={1},
  pages={24},
  year={2019},
  publisher={Nature Publishing Group UK London}
}

@article{ref11,
  title={Artificial intelligence in retina},
  author={Schmidt-Erfurth, Ursula and Sadeghipour, Amir and Gerendas, Bianca S and Waldstein, Sebastian M and Bogunovi{\'c}, Hrvoje},
  journal={Progress in retinal and eye research},
  volume={67},
  pages={1--29},
  year={2018},
  publisher={Elsevier}
}

@article{ref12,
  title={Physics-informed deep generative learning for quantitative assessment of the retina},
  author={Brown, Emmeline E and Guy, Andrew A and Holroyd, Natalie A and Sweeney, Paul W and Gourmet, Lucie and Coleman, Hannah and Walsh, Claire and Markaki, Athina E and Shipley, Rebecca and Rajendram, Ranjan and others},
  journal={Nature Communications},
  volume={15},
  number={1},
  pages={6859},
  year={2024},
  publisher={Nature Publishing Group UK London}
}

@article{ref13,
  title={A foundation model for generalizable disease detection from retinal images},
  author={Zhou, Yukun and Chia, Mark A and Wagner, Siegfried K and Ayhan, Murat S and Williamson, Dominic J and Struyven, Robbert R and Liu, Timing and Xu, Moucheng and Lozano, Mateo G and Woodward-Court, Peter and others},
  journal={Nature},
  volume={622},
  number={7981},
  pages={156--163},
  year={2023},
  publisher={Nature Publishing Group UK London}
}

@article{ref14,
  title={Clinically applicable deep learning for diagnosis and referral in retinal disease},
  author={De Fauw, Jeffrey and Ledsam, Joseph R and Romera-Paredes, Bernardino and Nikolov, Stanislav and Tomasev, Nenad and Blackwell, Sam and Askham, Harry and Glorot, Xavier and O’Donoghue, Brendan and Visentin, Daniel and others},
  journal={Nature medicine},
  volume={24},
  number={9},
  pages={1342--1350},
  year={2018},
  publisher={Nature Publishing Group US New York}
}

@article{ref15,
  title={Prediction of cardiovascular risk factors from retinal fundus photographs via deep learning},
  author={Poplin, Ryan and Varadarajan, Avinash V and Blumer, Katy and Liu, Yun and McConnell, Michael V and Corrado, Greg S and Peng, Lily and Webster, Dale R},
  journal={Nature biomedical engineering},
  volume={2},
  number={3},
  pages={158--164},
  year={2018},
  publisher={Nature Publishing Group UK London}
}

@article{ref16,
  title={Automated analysis of retinal imaging using machine learning techniques for computer vision},
  author={De Fauw, Jeffrey and Keane, Pearse and Tomasev, Nenad and Visentin, Daniel and van den Driessche, George and Johnson, Mike and Hughes, Cian O and Chu, Carlton and Ledsam, Joseph and Back, Trevor and others},
  journal={F1000Research},
  volume={5},
  pages={1573},
  year={2017}
}

@article{ref17,
  title={What makes UK Biobank special?},
  author={Collins, Rory},
  journal={Lancet (London, England)},
  volume={379},
  number={9822},
  pages={1173--1174},
  year={2012}
}

@article{ref18,
  title={10 K: a large-scale prospective longitudinal study in Israel},
  author={Shilo, Smadar and Bar, Noam and Keshet, Ayya and Talmor-Barkan, Yeela and Rossman, Hagai and Godneva, Anastasia and Aviv, Yaron and Edlitz, Yochai and Reicher, Lee and Kolobkov, Dmitry and others},
  journal={European journal of epidemiology},
  volume={36},
  number={11},
  pages={1187--1194},
  year={2021},
  publisher={Springer}
}

@article{ref19,
  title={A global review of publicly available datasets for ophthalmological imaging: barriers to access, usability, and generalisability},
  author={Khan, Saad M and Liu, Xiaoxuan and Nath, Siddharth and Korot, Edward and Faes, Livia and Wagner, Siegfried K and Keane, Pearse A and Sebire, Neil J and Burton, Matthew J and Denniston, Alastair K},
  journal={The Lancet Digital Health},
  volume={3},
  number={1},
  pages={e51--e66},
  year={2021},
  publisher={Elsevier}
}

@incollection{ref20,
  title={Artificial Intelligence and Machine Learning for Analysis of Multi-omics},
  author={Agarwal, Neeraj and Nupur and Paul, Prabir Kumar and Mishra, Santosh Kumar},
  booktitle={Multi-Omics Analysis of the Human Microbiome: From Technology to Clinical Applications},
  pages={339--354},
  year={2024},
  publisher={Springer}
}

@article{ref21,
  title={Methodological and statistical considerations for cross-sectional, case--control, and cohort studies},
  author={P{\'e}rez-Guerrero, Edsa{\'u}l Emilio and Guill{\'e}n-Medina, Miryam Rosario and M{\'a}rquez-Sandoval, Fabiola and Vera-Cruz, Jos{\'e} Mar{\'\i}a and Gallegos-Arreola, Martha Patricia and Rico-M{\'e}ndez, Manuel Alejandro and Aguilar-Vel{\'a}zquez, Jos{\'e} Alonso and Guti{\'e}rrez-Hurtado, Itzae Adonai},
  journal={Journal of Clinical Medicine},
  volume={13},
  number={14},
  pages={4005},
  year={2024},
  publisher={MDPI}
}

@article{ref22,
  title={Global burden of 369 diseases and injuries in 204 countries and territories, 1990--2019: a systematic analysis for the Global Burden of Disease Study 2019},
  author={Vos, Theo and Lim, Stephen S and Abbafati, Cristiana and Abbas, Kaja M and Abbasi, Mohammad and Abbasifard, Mitra and Abbasi-Kangevari, Mohsen and Abbastabar, Hedayat and Abd-Allah, Foad and Abdelalim, Ahmed and others},
  journal={The lancet},
  volume={396},
  number={10258},
  pages={1204--1222},
  year={2020},
  publisher={Elsevier}
}

@article{ref23,
  title={Recent Advances in Data-driven Fusion of Multi-modal Imaging and Genomics for Precision Medicine},
  author={Wang, Shuo and Liu, Meng and Li, Yan and Zhang, Xinyu and Sun, Mengting and Wang, Zian and Li, Ruokun and Li, Qirong and Li, Qing and He, Yili and others},
  journal={Information Fusion},
  pages={102738},
  year={2024},
  publisher={Elsevier}
}

@article{ref24,
  title={Multi-modal retinal scanning to measure retinal thickness and peripheral blood vessels in multiple sclerosis},
  author={Pearson, Thomas and Chen, Yingdi and Dhillon, Baljean and Chandran, Siddharthan and van Hemert, Jano and MacGillivray, Tom},
  journal={Scientific Reports},
  volume={12},
  number={1},
  pages={20472},
  year={2022},
  publisher={Nature Publishing Group UK London}
}

@incollection{ref25,
  title={Artificial intelligence in ophthalmology III: systemic disease prediction},
  author={Ran, An Ran and Hui, Herbert YH and Cheung, Carol Y and Wong, Tien Yin},
  booktitle={Artificial Intelligence in Clinical Practice},
  pages={119--125},
  year={2024},
  publisher={Elsevier}
}

@article{ref26,
  title={Artificial intelligence and deep learning in ophthalmology: current status and future perspectives},
  author={Jin, Kai and Ye, Juan},
  journal={Advances in ophthalmology practice and research},
  volume={2},
  number={3},
  pages={100078},
  year={2022},
  publisher={Elsevier}
}

@article{ref27,
  title={Artificial intelligence in retinal screening using OCT images: A review of the last decade (2013--2023)},
  author={Akpinar, Muhammed Halil and Sengur, Abdulkadir and Faust, Oliver and Tong, Louis and Molinari, Filippo and Acharya, U Rajendra},
  journal={Computer methods and programs in biomedicine},
  volume={254},
  pages={108253},
  year={2024},
  publisher={Elsevier}
}

@article{ref28,
  title={Association between retinal vessels caliber and systemic health: A comprehensive review},
  author={Lee, Si Jin Vanessa and Goh, Ying Qi and Rojas-Carabali, William and Cifuentes-Gonz{\'a}lez, Carlos and Cheung, Carol Y and Arora, Atul and de-la-Torre, Alejandra and Gupta, Vishali and Agrawal, Rupesh and others},
  journal={Survey of Ophthalmology},
  year={2024},
  publisher={Elsevier}
}

@article{ref29,
  title={Transformative applications of oculomics-based AI approaches in the management of systemic diseases: A systematic review},
  author={Li, Zhongwen and Yin, Shiqi and Wang, Shihong and Wang, Yangyang and Qiang, Wei and Jiang, Jiewei},
  journal={Journal of Advanced Research},
  year={2024},
  publisher={Elsevier}
}

@article{ref29a,
  title={Diagnostic patterns in keratoconus},
  author={Kreps, Elke O and Claerhout, Ilse and Koppen, Carina},
  journal={Contact Lens and Anterior Eye},
  volume={44},
  number={3},
  pages={101333},
  year={2021},
  publisher={Elsevier}
}

@article{ref29b,
  title={Myopia: anatomic changes and consequences for its etiology},
  author={Jonas, Jost B and Ohno-Matsui, Kyoko and Panda-Jonas, Songhomitra},
  journal={Asia-Pacific Journal of Ophthalmology},
  volume={8},
  number={5},
  pages={355--359},
  year={2019},
  publisher={Elsevier}
}

@article{ref29c,
  title={Hallmarks of lens aging and cataractogenesis},
  author={Wishart, Tayler FL and Flokis, Mary and Shu, Daisy Y and Das, Shannon J and Lovicu, Frank J},
  journal={Experimental eye research},
  volume={210},
  pages={108709},
  year={2021},
  publisher={Elsevier}
}

@incollection{ref30,
  title={Eye: Anatomy, physiology and barriers to drug delivery},
  author={Cholkar, Kishore and Dasari, Supriya Reddy and Pal, Dhananjay and Mitra, Ashim K},
  booktitle={Ocular transporters and receptors},
  pages={1--36},
  year={2013},
  publisher={Elsevier}
}

@article{ref31,
  title={A multi-omics atlas of the human retina at single-cell resolution},
  author={Liang, Qingnan and Cheng, Xuesen and Wang, Jun and Owen, Leah and Shakoor, Akbar and Lillvis, John L and Zhang, Charles and Farkas, Michael and Kim, Ivana K and Li, Yumei and others},
  journal={Cell genomics},
  volume={3},
  number={6},
  year={2023},
  publisher={Elsevier}
}

@article{ref32,
  title={Retinal imaging and image analysis},
  author={Abr{\`a}moff, Michael D and Garvin, Mona K and Sonka, Milan},
  journal={IEEE reviews in biomedical engineering},
  volume={3},
  pages={169--208},
  year={2010},
  publisher={IEEE}
}

@article{ref32a,
  title={Retinal vascular caliber as a biomarker for diabetes microvascular complications},
  author={Ikram, M Kamran and Cheung, Carol Y and Lorenzi, Mara and Klein, Ronald and Jones, Teresa LZ and Wong, Tien Yin and others},
  journal={Diabetes care},
  volume={36},
  number={3},
  pages={750},
  year={2013}
}

@misc{ref32b,
  title={Retinal imaging in Alzheimer’s and neurodegenerative diseases. Alzheimers Dement 17 (1): 103--111},
  author={Snyder, PJ and Alber, J and Alt, C and Bain, LJ and Bouma, BE and Bouwman, FH and DeBuc, DC and Campbell, MCW and Carrillo, MC and Chew, EY and others},
  year={2021}
}

@article{ref33,
  title={Couching for cataract in China},
  author={Chan, Chi-Chao},
  journal={Survey of ophthalmology},
  volume={55},
  number={4},
  pages={393--398},
  year={2010},
  publisher={Elsevier}
}

@article{ref34,
  title={The history of uveitis: from antiquity to the present day},
  author={Ghadiri, Nima},
  journal={Eye},
  pages={1--4},
  year={2024},
  publisher={Nature Publishing Group}
}

@article{ref35,
  title={A history of the optic nerve and its diseases},
  author={Reeves, Carole and Taylor, David},
  journal={Eye},
  volume={18},
  number={11},
  pages={1096--1109},
  year={2004},
  publisher={Nature Publishing Group}
}

@article{ref36,
  title={The evolution of the ophthalmoscope},
  author={McMullen, WH},
  journal={The British Journal of Ophthalmology},
  volume={1},
  number={10},
  pages={593},
  year={1917},
  publisher={BMJ Publishing Group}
}

@article{ref37,
  title={Infectious eye disease in the 21st century—an overview},
  author={Clare, Gerry and Kempen, John H and Pav{\'e}sio, Carlos},
  journal={Eye},
  pages={1--14},
  year={2024},
  publisher={Nature Publishing Group UK London}
}

@incollection{ref38,
  title={Fluorescein Angiography},
  author={Weiss, Stephanie J and Papakostas, Thanos D},
  booktitle={Albert and Jakobiec's Principles and Practice of Ophthalmology},
  pages={2659--2682},
  year={2022},
  publisher={Springer}
}

@article{ref39,
  title={Different lasers and techniques for proliferative diabetic retinopathy},
  author={Moutray, Tanya and Evans, Jennifer R and Lois, Noemi and Armstrong, David J and Peto, Tunde and Azuara-Blanco, Augusto},
  journal={Cochrane Database of Systematic Reviews},
  number={3},
  year={2018},
  publisher={John Wiley \& Sons, Ltd}
}

@article{ref40,
  title={Tonometry and Intraocular Pressure--Where are we Now?},
  author={Garc{\'\i}a-Feijoo, Julian},
  journal={Journal-Tonometry and Intraocular Pressure--Where are we Now?}
}

@article{ref41,
  title={Optical coherence tomography (OCT): principle and technical realization},
  author={Aumann, Silke and Donner, Sabine and Fischer, J{\"o}rg and M{\"u}ller, Frank},
  journal={High resolution imaging in microscopy and ophthalmology: new frontiers in biomedical optics},
  pages={59--85},
  year={2019},
  publisher={Springer}
}

@ARTICLE{ref42,
  author={Remeseiro, Beatriz and Bolon-Canedo, Veronica and Peteiro-Barral, Diego and Alonso-Betanzos, Amparo and Guijarro-Berdiñas, Bertha and Mosquera, Antonio and Penedo, Manuel G. and Sánchez-Maroño, Noelia},
  journal={IEEE Journal of Biomedical and Health Informatics}, 
  title={A Methodology for Improving Tear Film Lipid Layer Classification}, 
  year={2014},
  volume={18},
  number={4},
  pages={1485-1493},
  doi={10.1109/JBHI.2013.2294732}}

@article{ref43,
  title={Automated macular pathology diagnosis in retinal OCT images using multi-scale spatial pyramid and local binary patterns in texture and shape encoding},
  author={Liu, Yu-Ying and Chen, Mei and Ishikawa, Hiroshi and Wollstein, Gadi and Schuman, Joel S and Rehg, James M},
  journal={Medical image analysis},
  volume={15},
  number={5},
  pages={748--759},
  year={2011},
  publisher={Elsevier}
}

@article{ref44,
  title={Automatic microaneurysm detection using laws texture masks and support vector machines},
  author={Veiga, Diana and Martins, Nelson and Ferreira, Manuel and Monteiro, Jo{\~a}o},
  journal={Computer Methods in Biomechanics and Biomedical Engineering: Imaging \& Visualization},
  volume={6},
  number={4},
  pages={405--416},
  year={2018},
  publisher={Taylor \& Francis}
}

@article{ref45,
  title={Macula segmentation and fovea localization employing image processing and heuristic based clustering for automated retinal screening},
  author={GeethaRamani, R and Balasubramanian, Lakshmi},
  journal={Computer methods and programs in biomedicine},
  volume={160},
  pages={153--163},
  year={2018},
  publisher={Elsevier}
}

@article{ref46,
  title={Development and validation of a deep learning algorithm for detection of diabetic retinopathy in retinal fundus photographs},
  author={Gulshan, Varun and Peng, Lily and Coram, Marc and Stumpe, Martin C and Wu, Derek and Narayanaswamy, Arunachalam and Venugopalan, Subhashini and Widner, Kasumi and Madams, Tom and Cuadros, Jorge and others},
  journal={jama},
  volume={316},
  number={22},
  pages={2402--2410},
  year={2016},
  publisher={American Medical Association}
}

@article{ref47,
  title={Development and validation of a deep learning system for diabetic retinopathy and related eye diseases using retinal images from multiethnic populations with diabetes},
  author={Ting, Daniel Shu Wei and Cheung, Carol Yim-Lui and Lim, Gilbert and Tan, Gavin Siew Wei and Quang, Nguyen D and Gan, Alfred and Hamzah, Haslina and Garcia-Franco, Renata and San Yeo, Ian Yew and Lee, Shu Yen and others},
  journal={Jama},
  volume={318},
  number={22},
  pages={2211--2223},
  year={2017},
  publisher={American Medical Association}
}

@misc{ref48,
      title={An Image is Worth 16x16 Words: Transformers for Image Recognition at Scale}, 
      author={Alexey Dosovitskiy and Lucas Beyer and Alexander Kolesnikov and Dirk Weissenborn and Xiaohua Zhai and Thomas Unterthiner and Mostafa Dehghani and Matthias Minderer and Georg Heigold and Sylvain Gelly and Jakob Uszkoreit and Neil Houlsby},
      year={2021},
      eprint={2010.11929},
      archivePrefix={arXiv},
      primaryClass={cs.CV},
      url={https://arxiv.org/abs/2010.11929}, 
}

@inproceedings{ref49,
  title={Mil-vt: Multiple instance learning enhanced vision transformer for fundus image classification},
  author={Yu, Shuang and Ma, Kai and Bi, Qi and Bian, Cheng and Ning, Munan and He, Nanjun and Li, Yuexiang and Liu, Hanruo and Zheng, Yefeng},
  booktitle={Medical Image Computing and Computer Assisted Intervention--MICCAI 2021: 24th International Conference, Strasbourg, France, September 27--October 1, 2021, Proceedings, Part VIII 24},
  pages={45--54},
  year={2021},
  organization={Springer}
}

@article{ref50,
  title={Deep learning of the retina enables phenome-and genome-wide analyses of the microvasculature},
  author={Zekavat, Seyedeh Maryam and Raghu, Vineet K and Trinder, Mark and Ye, Yixuan and Koyama, Satoshi and Honigberg, Michael C and Yu, Zhi and Pampana, Akhil and Urbut, Sarah and Haidermota, Sara and others},
  journal={Circulation},
  volume={145},
  number={2},
  pages={134--150},
  year={2022},
  publisher={Am Heart Assoc}
}

@article{ref51,
  title={Multi-omics in exploring the pathophysiology of diabetic retinopathy},
  author={Li, Xinlu and Dong, XiaoJing and Zhang, Wen and Shi, Zhizhou and Liu, Zhongjian and Sa, Yalian and Li, Li and Ni, Ninghua and Mei, Yan},
  journal={Frontiers in Cell and Developmental Biology},
  volume={12},
  pages={1500474},
  year={2024},
  publisher={Frontiers Media SA}
}

@article{ref52,
  title={A large genome-wide association study of age-related macular degeneration highlights contributions of rare and common variants},
  author={Fritsche, Lars G and Igl, Wilmar and Bailey, Jessica N Cooke and Grassmann, Felix and Sengupta, Sebanti and Bragg-Gresham, Jennifer L and Burdon, Kathryn P and Hebbring, Scott J and Wen, Cindy and Gorski, Mathias and others},
  journal={Nature genetics},
  volume={48},
  number={2},
  pages={134--143},
  year={2016},
  publisher={Nature Publishing Group US New York}
}

@article{ref53,
  title={Genome-wide meta-analysis identifies 127 open-angle glaucoma loci with consistent effect across ancestries},
  author={Gharahkhani, Puya and Jorgenson, Eric and Hysi, Pirro and Khawaja, Anthony P and Pendergrass, Sarah and Han, Xikun and Ong, Jue Sheng and Hewitt, Alex W and Segr{\`e}, Ayellet V and Rouhana, John M and others},
  journal={Nature communications},
  volume={12},
  number={1},
  pages={1258},
  year={2021},
  publisher={Nature Publishing Group UK London}
}

@article{ref54,
  title={Cross-species scRNA-seq reveals the cellular landscape of retina and early alterations in type 2 diabetes mice},
  author={Chen, Kai and Wang, Yinhao and Huang, Youyuan and Liu, Xinxin and Tian, Xiaodong and Yang, Yinmo and Dong, Aimei},
  journal={Genomics},
  volume={115},
  number={4},
  pages={110644},
  year={2023},
  publisher={Elsevier}
}

@article{ref55,
  title={Genome editing in the treatment of ocular diseases},
  author={Choi, Elliot H and Suh, Susie and Sears, Avery E and Ho{\l}ubowicz, Rafa{\l} and Kedhar, Sanjay R and Browne, Andrew W and Palczewski, Krzysztof},
  journal={Experimental \& molecular medicine},
  volume={55},
  number={8},
  pages={1678--1690},
  year={2023},
  publisher={Nature Publishing Group UK London}
}

@article{ref56,
  title={A survey of large language models for healthcare: from data, technology, and applications to accountability and ethics},
  author={He, Kai and Mao, Rui and Lin, Qika and Ruan, Yucheng and Lan, Xiang and Feng, Mengling and Cambria, Erik},
  journal={Information Fusion},
  pages={102963},
  year={2025},
  publisher={Elsevier}
}

@article{ref57,
  title={Evaluating the performance of ChatGPT in ophthalmology: an analysis of its successes and shortcomings},
  author={Antaki, Fares and Touma, Samir and Milad, Daniel and El-Khoury, Jonathan and Duval, Renaud},
  journal={Ophthalmology science},
  volume={3},
  number={4},
  pages={100324},
  year={2023},
  publisher={Elsevier}
}

@article{ref58,
  title={Large language models and their impact in ophthalmology},
  author={Betzler, Bjorn Kaijun and Chen, Haichao and Cheng, Ching-Yu and Lee, Cecilia S and Ning, Guochen and Song, Su Jeong and Lee, Aaron Y and Kawasaki, Ryo and van Wijngaarden, Peter and Grzybowski, Andrzej and others},
  journal={The Lancet Digital Health},
  volume={5},
  number={12},
  pages={e917--e924},
  year={2023},
  publisher={Elsevier}
}

@article{ref59,
  title={A guide to artificial intelligence for cancer researchers},
  author={Perez-Lopez, Raquel and Ghaffari Laleh, Narmin and Mahmood, Faisal and Kather, Jakob Nikolas},
  journal={Nature Reviews Cancer},
  pages={1--15},
  year={2024},
  publisher={Nature Publishing Group UK London}
}

@article{ref60,
  title={Deep learning},
  author={LeCun, Yann and Bengio, Yoshua and Hinton, Geoffrey},
  journal={nature},
  volume={521},
  number={7553},
  pages={436--444},
  year={2015},
  publisher={Nature Publishing Group UK London}
}

@article{ref61,
  title={Deep convolutional neural networks for image classification: A comprehensive review},
  author={Rawat, Waseem and Wang, Zenghui},
  journal={Neural computation},
  volume={29},
  number={9},
  pages={2352--2449},
  year={2017},
  publisher={MIT Press}
}

@article{ref61a,
  title={Optical coherence tomography and glaucoma},
  author={Geevarghese, Alexi and Wollstein, Gadi and Ishikawa, Hiroshi and Schuman, Joel S},
  journal={Annual review of vision science},
  volume={7},
  number={1},
  pages={693--726},
  year={2021},
  publisher={Annual Reviews}
}

@article{ref61b,
  title={SD-OCT peripapillary nerve fibre layer and ganglion cell complex parameters in glaucoma: principal component analysis},
  author={Pazos, Marta and Biarn{\'e}s, Marc and Blasco-Alberto, Andr{\'e}s and Dyrda, Agnieszka and Luque-Fern{\'a}ndez, Miguel {\'A}ngel and G{\'o}mez, Alicia and Mora, Clara and Milla, Elena and Muniesa, M{\textordfeminine}Jes{\'u}s and Ant{\'o}n, Alfonso and others},
  journal={British Journal of Ophthalmology},
  volume={105},
  number={4},
  pages={496--501},
  year={2021},
  publisher={BMJ Publishing Group Ltd}
}

@incollection{ref61c,
  title={Fluorescein angiography},
  author={Ricardi, Federico and Reibaldi, Michele and Bandello, Francesco and Borrelli, Enrico},
  booktitle={Retinal and Choroidal Vascular Diseases of the Eye},
  pages={71--79},
  year={2024},
  publisher={Elsevier}
}

@article{ref62,
  title={A practical guide to optical coherence tomography angiography interpretation},
  author={Greig, Eugenia Custo and Duker, Jay S and Waheed, Nadia K},
  journal={International journal of retina and vitreous},
  volume={6},
  pages={1--17},
  year={2020},
  publisher={Springer}
}

@article{ref63,
  title={The new era of retinal imaging in hypertensive patients},
  author={Tan, Wilson and Yao, Xinwen and Le, Thu-Thao and Tan, Bingyao and Schmetterer, Leopold and Chua, Jacqueline},
  journal={The Asia-Pacific Journal of Ophthalmology},
  volume={11},
  number={2},
  pages={149--159},
  year={2022},
  publisher={LWW}
}

@article{ref63a,
  title={Concordance between SIVA, IVAN, and VAMPIRE software tools for semi-automated analysis of retinal vessel caliber},
  author={Mautuit, Thibaud and Cunnac, Pierre and Cheung, Carol Y and Wong, Tien Y and Hogg, Stephen and Trucco, Emanuele and Daien, Vincent and MacGillivray, Thomas J and Labar{\`e}re, Jos{\'e} and Chiquet, Christophe},
  journal={Diagnostics},
  volume={12},
  number={6},
  pages={1317},
  year={2022},
  publisher={MDPI}
}

@article{ref63b,
  title={Wide-field imaging of the retina},
  author={Witmer, Matthew T and Kiss, Szil{\'a}rd},
  journal={Survey of ophthalmology},
  volume={58},
  number={2},
  pages={143--154},
  year={2013},
  publisher={Elsevier}
}

@article{ref63c,
  title={Retinopathy of prematurity detection: a retrospective quality improvement project before-after implementation of retinal digital imaging for screening},
  author={Desurmont, Marie-Gwenola and Bremond-Gignac, Dominique and Torchin, H{\'e}lo{\"\i}se and Vacherot, Brigitte and Jarreau, Pierre-Henri and Daruich, Alejandra},
  journal={European Journal of Pediatrics},
  volume={182},
  number={7},
  pages={3093--3099},
  year={2023},
  publisher={Springer}
}

@article{ref63d,
  title={Non-invasive in vivo hyperspectral imaging of the retina for potential biomarker use in Alzheimer’s disease},
  author={Hadoux, Xavier and Hui, Flora and Lim, Jeremiah KH and Masters, Colin L and P{\'e}bay, Alice and Chevalier, Sophie and Ha, Jason and Loi, Samantha and Fowler, Christopher J and Rowe, Christopher and others},
  journal={Nature communications},
  volume={10},
  number={1},
  pages={4227},
  year={2019},
  publisher={Nature Publishing Group UK London}
}

@article{ref63e,
  title={Adaptive optics imaging of inherited retinal diseases},
  author={Georgiou, Michalis and Kalitzeos, Angelos and Patterson, Emily J and Dubra, Alfredo and Carroll, Joseph and Michaelides, Michel},
  journal={British Journal of Ophthalmology},
  volume={102},
  number={8},
  pages={1028--1035},
  year={2018},
  publisher={BMJ Publishing Group Ltd}
}

@article{ref63f,
  title={Is retinal photography useful in the measurement of stroke risk?},
  author={Wong, Tien Yin},
  journal={The Lancet Neurology},
  volume={3},
  number={3},
  pages={179--183},
  year={2004},
  publisher={Elsevier}
}

@article{ref63g,
  title={Blue widefield images of scanning laser ophthalmoscope can detect retinal ischemic areas in eyes with diabetic retinopathy},
  author={Horie, Shintaro and Kukimoto, Nobuyuki and Kamoi, Koju and Igarashi-Yokoi, Tae and Yoshida, Takeshi and Ohno-Matsui, Kyoko},
  journal={Asia-Pacific Journal of Ophthalmology},
  volume={10},
  number={5},
  pages={478--485},
  year={2021},
  publisher={Elsevier}
}

@article{ref65,
  title={Applications of deep learning in fundus images: A review},
  author={Li, Tao and Bo, Wang and Hu, Chunyu and Kang, Hong and Liu, Hanruo and Wang, Kai and Fu, Huazhu},
  journal={Medical Image Analysis},
  volume={69},
  pages={101971},
  year={2021},
  publisher={Elsevier}
}

@article{ref66,
  title={Locating blood vessels in retinal images by piecewise threshold probing of a matched filter response},
  author={Hoover, AD and Kouznetsova, Valentina and Goldbaum, Michael},
  journal={IEEE Transactions on Medical imaging},
  volume={19},
  number={3},
  pages={203--210},
  year={2000},
  publisher={IEEE}
}

@article{ref67,
  title={Ridge-based vessel segmentation in color images of the retina},
  author={Staal, Joes and Abr{\`a}moff, Michael D and Niemeijer, Meindert and Viergever, Max A and Van Ginneken, Bram},
  journal={IEEE transactions on medical imaging},
  volume={23},
  number={4},
  pages={501--509},
  year={2004},
  publisher={IEEE}
}

@inproceedings{ref68,
  title={The diaretdb1 diabetic retinopathy database and evaluation protocol.},
  author={Kauppi, Tomi and Kalesnykiene, Valentina and Kamarainen, Joni-Kristian and Lensu, Lasse and Sorri, Iiris and Raninen, Asta and Voutilainen, Raija and Uusitalo, Hannu and K{\"a}lvi{\"a}inen, Heikki and Pietil{\"a}, Juhani},
  booktitle={BMVC},
  volume={1},
  number={1},
  pages={10},
  year={2007},
  organization={Citeseer}
}

@article{ref69,
  title={Measuring retinal vessel tortuosity in 10-year-old children: validation of the computer-assisted image analysis of the retina (CAIAR) program},
  author={Owen, Christopher G and Rudnicka, Alicja R and Mullen, Robert and Barman, Sarah A and Monekosso, Dorothy and Whincup, Peter H and Ng, Jeffrey and Paterson, Carl},
  journal={Investigative ophthalmology \& visual science},
  volume={50},
  number={5},
  pages={2004--2010},
  year={2009},
  publisher={The Association for Research in Vision and Ophthalmology}
}

@article{ref70,
  title={Robust vessel segmentation in fundus images},
  author={Budai, Attila and Bock, R{\"u}diger and Maier, Andreas and Hornegger, Joachim and Michelson, Georg},
  journal={International journal of biomedical imaging},
  volume={2013},
  number={1},
  pages={154860},
  year={2013},
  publisher={Wiley Online Library}
}

@article{ref71,
  title={TeleOphta: Machine learning and image processing methods for teleophthalmology},
  author={Decenciere, Etienne and Cazuguel, Guy and Zhang, Xiwei and Thibault, Guillaume and Klein, J-C and Meyer, Fernand and Marcotegui, Beatriz and Quellec, Gw{\'e}nol{\'e} and Lamard, Mathieu and Danno, Ronan and others},
  journal={Irbm},
  volume={34},
  number={2},
  pages={196--203},
  year={2013},
  publisher={Elsevier}
}

@article{ref72,
  title={Feedback on a publicly distributed image database: the Messidor database},
  author={Decenci{\`e}re, Etienne and Zhang, Xiwei and Cazuguel, Guy and Lay, Bruno and Cochener, B{\'e}atrice and Trone, Caroline and Gain, Philippe and Ord{\'o}{\~n}ez-Varela, John-Richard and Massin, Pascale and Erginay, Ali and others},
  journal={Image Analysis \& Stereology},
  pages={231--234},
  year={2014}
}

@misc{ref73,
    author = {Emma Dugas and Jared and Jorge and Will Cukierski},
    title = {Diabetic Retinopathy Detection},
    year = {2015},
    howpublished = {\url{https://kaggle.com/competitions/diabetic-retinopathy-detection}},
    note = {Kaggle}
}

@article{ref74,
  title={Indian diabetic retinopathy image dataset (IDRiD): a database for diabetic retinopathy screening research},
  author={Porwal, Prasanna and Pachade, Samiksha and Kamble, Ravi and Kokare, Manesh and Deshmukh, Girish and Sahasrabuddhe, Vivek and Meriaudeau, Fabrice},
  journal={Data},
  volume={3},
  number={3},
  pages={25},
  year={2018},
  publisher={MDPI}
}

@misc{ref75,
    author = {Karthik and Maggie and Sohier Dane},
    title = {APTOS 2019 Blindness Detection},
    year = {2019},
    howpublished = {\url{https://kaggle.com/competitions/aptos2019-blindness-detection}},
    note = {Kaggle}
}

@article{ref76,
  title={Diagnostic assessment of deep learning algorithms for diabetic retinopathy screening},
  author={Li, Tao and Gao, Yingqi and Wang, Kai and Guo, Song and Liu, Hanruo and Kang, Hong},
  journal={Information Sciences},
  volume={501},
  pages={511--522},
  year={2019},
  publisher={Elsevier}
}

@article{ref77,
  title={Global prevalence of glaucoma and projections of glaucoma burden through 2040: a systematic review and meta-analysis},
  author={Tham, Yih-Chung and Li, Xiang and Wong, Tien Y and Quigley, Harry A and Aung, Tin and Cheng, Ching-Yu},
  journal={Ophthalmology},
  volume={121},
  number={11},
  pages={2081--2090},
  year={2014},
  publisher={Elsevier}
}

@article{ref78,
  title={Updates on the Diagnosis and Management of Glaucoma},
  author={Wagner, Isabella V and Stewart, Michael W and Dorairaj, Syril K},
  journal={Mayo Clinic Proceedings: Innovations, Quality \& Outcomes},
  volume={6},
  number={6},
  pages={618--635},
  year={2022},
  publisher={Elsevier}
}

@article{ref79,
  title={A review on glaucoma: causes, symptoms, pathogenesis \& treatment},
  author={Sahu, Mahendra Kumar},
  journal={Journal of Clinical Research and Ophthalmology},
  volume={11},
  number={1},
  pages={001--004},
  year={2024}
}

@article{ref80,
  title={Automatic detection of glaucoma via fundus imaging and artificial intelligence: A review},
  author={Coan, Lauren J and Williams, Bryan M and Adithya, Venkatesh Krishna and Upadhyaya, Swati and Alkafri, Ala and Czanner, Silvester and Venkatesh, Rengaraj and Willoughby, Colin E and Kavitha, Srinivasan and Czanner, Gabriela},
  journal={Survey of ophthalmology},
  volume={68},
  number={1},
  pages={17--41},
  year={2023},
  publisher={Elsevier}
}

@article{ref81,
  title={Automated detection of glaucoma with interpretable machine learning using clinical data and multimodal retinal images},
  author={Mehta, Parmita and Petersen, Christine A and Wen, Joanne C and Banitt, Michael R and Chen, Philip P and Bojikian, Karine D and Egan, Catherine and Lee, Su-In and Balazinska, Magdalena and Lee, Aaron Y and others},
  journal={American Journal of Ophthalmology},
  volume={231},
  pages={154--169},
  year={2021},
  publisher={Elsevier}
}

@article{ref82,
  title={Glaucoma risk index: automated glaucoma detection from color fundus images},
  author={Bock, R{\"u}diger and Meier, J{\"o}rg and Ny{\'u}l, L{\'a}szl{\'o} G and Hornegger, Joachim and Michelson, Georg},
  journal={Medical image analysis},
  volume={14},
  number={3},
  pages={471--481},
  year={2010},
  publisher={Elsevier}
}

@article{ref83,
  title={Efficacy of a deep learning system for detecting glaucomatous optic neuropathy based on color fundus photographs},
  author={Li, Zhixi and He, Yifan and Keel, Stuart and Meng, Wei and Chang, Robert T and He, Mingguang},
  journal={Ophthalmology},
  volume={125},
  number={8},
  pages={1199--1206},
  year={2018},
  publisher={Elsevier}
}

@article{ref84,
  title={Policy-driven, multimodal deep learning for predicting visual fields from the optic disc and OCT imaging},
  author={Kihara, Yuka and Montesano, Giovanni and Chen, Andrew and Amerasinghe, Nishani and Dimitriou, Chrysostomos and Jacob, Aby and Chabi, Almira and Crabb, David P and Lee, Aaron Y},
  journal={Ophthalmology},
  volume={129},
  number={7},
  pages={781--791},
  year={2022},
  publisher={Elsevier}
}

@article{ref85,
  title={Improving visual field trend analysis with OCT and deeply regularized latent-space linear regression},
  author={Xu, Linchuan and Asaoka, Ryo and Murata, Hiroshi and Kiwaki, Taichi and Zheng, Yuhui and Matsuura, Masato and Fujino, Yuri and Tanito, Masaki and Mori, Kazuhiko and Ikeda, Yoko and others},
  journal={Ophthalmology Glaucoma},
  volume={4},
  number={1},
  pages={78--88},
  year={2021},
  publisher={Elsevier}
}

@article{ref86,
  title={Focused attention in transformers for interpretable classification of retinal images},
  author={Playout, Cl{\'e}ment and Duval, Renaud and Boucher, Marie Carole and Cheriet, Farida},
  journal={Medical Image Analysis},
  volume={82},
  pages={102608},
  year={2022},
  publisher={Elsevier}
}

@article{ref87,
  title={Recent advances in the application of artificial intelligence in age-related macular degeneration},
  author={Gao, Yundi and Xiong, Fen and Xiong, Jian and Chen, Zidan and Lin, Yucai and Xia, Xinjing and Yang, Yulan and Li, Guodong and Hu, Yunwei},
  journal={BMJ Open Ophthalmology},
  volume={9},
  number={1},
  year={2024},
  publisher={BMJ Publishing Group Ltd}
}

@article{ref88,
  title={Global prevalence of age-related macular degeneration and disease burden projection for 2020 and 2040: a systematic review and meta-analysis},
  author={Wong, Wan Ling and Su, Xinyi and Li, Xiang and Cheung, Chui Ming G and Klein, Ronald and Cheng, Ching-Yu and Wong, Tien Yin},
  journal={The Lancet Global Health},
  volume={2},
  number={2},
  pages={e106--e116},
  year={2014},
  publisher={Elsevier}
}

@article{ref89,
  title={Automatic drusen quantification and risk assessment of age-related macular degeneration on color fundus images},
  author={van Grinsven, Mark JJP and Lechanteur, Yara TE and van de Ven, Johannes PH and van Ginneken, Bram and Hoyng, Carel B and Theelen, Thomas and S{\'a}nchez, Clara I},
  journal={Investigative ophthalmology \& visual science},
  volume={54},
  number={4},
  pages={3019--3027},
  year={2013},
  publisher={The Association for Research in Vision and Ophthalmology}
}

@article{ref90,
  title={A machine learning approach to medical image classification: Detecting age-related macular degeneration in fundus images},
  author={Garc{\'\i}a-Floriano, Andr{\'e}s and Ferreira-Santiago, {\'A}ngel and Camacho-Nieto, Oscar and Y{\'a}{\~n}ez-M{\'a}rquez, Cornelio},
  journal={Computers \& Electrical Engineering},
  volume={75},
  pages={218--229},
  year={2019},
  publisher={Elsevier}
}

@article{ref91,
  title={A concentrated machine learning-based classification system for age-related macular degeneration (AMD) diagnosis using fundus images},
  author={Abd El-Khalek, Aya A and Balaha, Hossam Magdy and Alghamdi, Norah Saleh and Ghazal, Mohammed and Khalil, Abeer T and Abo-Elsoud, Mohy Eldin A and El-Baz, Ayman},
  journal={Scientific Reports},
  volume={14},
  number={1},
  pages={2434},
  year={2024},
  publisher={Nature Publishing Group UK London}
}

@article{ref92,
  title={Automatic detection and differential diagnosis of age-related macular degeneration from color fundus photographs using deep learning with hierarchical vision transformer},
  author={Xu, Ke and Huang, Shenghai and Yang, Zijian and Zhang, Yibo and Fang, Ye and Zheng, Gongwei and Lin, Bin and Zhou, Meng and Sun, Jie},
  journal={Computers in Biology and Medicine},
  volume={167},
  pages={107616},
  year={2023},
  publisher={Elsevier}
}

@article{ref93,
  title={Automatic detection of age-related macular degeneration based on deep learning and local outlier factor algorithm},
  author={He, Tingting and Zhou, Qiaoer and Zou, Yuanwen},
  journal={Diagnostics},
  volume={12},
  number={2},
  pages={532},
  year={2022},
  publisher={MDPI}
}

@article{ref94,
  title={The increasing rate of diabetes in Pakistan: A silent killer},
  author={Azeem, Saleha and Khan, Ubaid and Liaquat, Ayesha},
  journal={Annals of medicine and surgery},
  volume={79},
  year={2022},
  publisher={LWW}
}

@article{ref95,
  title={Automatic detection of diabetic eye disease through deep learning using fundus images: a survey},
  author={Sarki, Rubina and Ahmed, Khandakar and Wang, Hua and Zhang, Yanchun},
  journal={IEEE access},
  volume={8},
  pages={151133--151149},
  year={2020},
  publisher={IEEE}
}

@article{ref96,
  title={The pathophysiological mechanisms underlying diabetic retinopathy},
  author={Wei, Lindan and Sun, Xin and Fan, Chenxi and Li, Rongli and Zhou, Shuanglong and Yu, Hongsong},
  journal={Frontiers in Cell and Developmental Biology},
  volume={10},
  pages={963615},
  year={2022},
  publisher={Frontiers Media SA}
}

@ARTICLE{ref97,
  author={Khan, Zubair and Khan, Fiaz Gul and Khan, Ahmad and Rehman, Zia Ur and Shah, Sajid and Qummar, Sehrish and Ali, Farman and Pack, Sangheon},
  journal={IEEE Access}, 
  title={Diabetic Retinopathy Detection Using VGG-NIN a Deep Learning Architecture}, 
  year={2021},
  volume={9},
  number={},
  pages={61408-61416},
  doi={10.1109/ACCESS.2021.3074422}
}

@article{ref98,
  title={Classification of diabetic retinopathy: Past, present and future},
  author={Yang, Zhengwei and Tan, Tien-En and Shao, Yan and Wong, Tien Yin and Li, Xiaorong},
  journal={Frontiers in Endocrinology},
  volume={13},
  pages={1079217},
  year={2022},
  publisher={Frontiers Media SA}
}

@article{ref99,
  title={Diabetic retinopathy detection through novel tetragonal local octa patterns and extreme learning machines},
  author={Nazir, Tahira and Irtaza, Aun and Shabbir, Zain and Javed, Ali and Akram, Usman and Mahmood, Muhammad Tariq},
  journal={Artificial intelligence in medicine},
  volume={99},
  pages={101695},
  year={2019},
  publisher={Elsevier}
}

@article{ref100,
  title={Diabetic retinopathy classification based on multipath CNN and machine learning classifiers},
  author={Gayathri, S and Gopi, Varun P and Palanisamy, P},
  journal={Physical and engineering sciences in medicine},
  volume={44},
  number={3},
  pages={639--653},
  year={2021},
  publisher={Springer}
}

@article{ref101,
  title={Investigation of severity level of diabetic retinopathy using adaboost classifier algorithm},
  author={Washburn, Praveen Samuel and others},
  journal={Materials Today: Proceedings},
  volume={33},
  pages={3037--3042},
  year={2020},
  publisher={Elsevier}
}

@article{ref102,
  title={Deep convolutional neural network-based early automated detection of diabetic retinopathy using fundus image},
  author={Xu, Kele and Feng, Dawei and Mi, Haibo},
  journal={Molecules},
  volume={22},
  number={12},
  pages={2054},
  year={2017},
  publisher={MDPI}
}

@article{ref103,
  title={Classification of diabetic and normal fundus images using new deep learning method},
  author={Esfahani, Mehdi Torabian and Ghaderi, Mahsa and Kafiyeh, Raheleh},
  journal={Leonardo Electron. J. Pract. Technol},
  volume={17},
  number={32},
  pages={233--248},
  year={2018}
}

@article{ref104,
  title={Diabetic retinopathy detection using red lesion localization and convolutional neural networks},
  author={Zago, Gabriel Tozatto and Andre{\~a}o, Rodrigo Varej{\~a}o and Dorizzi, Bernadette and Salles, Evandro Ottoni Teatini},
  journal={Computers in biology and medicine},
  volume={116},
  pages={103537},
  year={2020},
  publisher={Elsevier}
}

@inproceedings{ref105,
  title={Mobile assisted diabetic retinopathy detection using deep neural network},
  author={Suriyal, Shorav and Druzgalski, Christopher and Gautam, Kumar},
  booktitle={2018 Global medical engineering physics exchanges/pan American Health Care Exchanges (GMEPE/PAHCE)},
  pages={1--4},
  year={2018},
  organization={IEEE}
}

@article{ref106,
  title={Screening for diabetic retinopathy using an automated diagnostic system based on deep learning: diagnostic accuracy assessment},
  author={R{\^e}go, S{\'\i}lvia and Dutra-Medeiros, Marco and Soares, Filipe and Monteiro-Soares, Matilde},
  journal={Ophthalmologica},
  volume={244},
  number={3},
  pages={250--257},
  year={2021},
  publisher={S. Karger AG}
}

@inproceedings{ref107,
  title={Diabetic retinopathy detection using MobileNetV2 architecture},
  author={Pamadi, Abhay M and Ravishankar, Ananya and Nithya, P Anu and Jahnavi, G and Kathavate, Sheela},
  booktitle={2022 International Conference on Smart Technologies and Systems for Next Generation Computing (ICSTSN)},
  pages={1--5},
  year={2022},
  organization={IEEE}
}

@inproceedings{ref108,
  title={Detection of diabetic retinopathy in retinal fundus images using densenet based deep learning model},
  author={Saranya, P and Devi, S Kiruthika and Bharanidharan, B},
  booktitle={2022 international mobile and embedded technology conference (MECON)},
  pages={268--272},
  year={2022},
  organization={IEEE}
}

@inproceedings{ref109,
  title={Diabetic retinopathy classification with pre-trained image enhancement model},
  author={Mudaser, Wahidullah and Padungweang, Praisan and Mongkolnam, Pornchai and Lavangnananda, Patcharaporn},
  booktitle={2021 IEEE 12th Annual Ubiquitous Computing, Electronics \& Mobile Communication Conference (UEMCON)},
  pages={0629--0632},
  year={2021},
  organization={IEEE}
}

@inproceedings{ref110,
  title={Classification of diabetic retinopathy based on hybrid neural network},
  author={Boral, Yash S and Thorat, Snehal S},
  booktitle={2021 5th International Conference on Computing Methodologies and Communication (ICCMC)},
  pages={1354--1358},
  year={2021},
  organization={IEEE}
}

@inproceedings{ref112,
  title={An interpretable ensemble deep learning model for diabetic retinopathy disease classification},
  author={Jiang, Hongyang and Yang, Kang and Gao, Mengdi and Zhang, Dongdong and Ma, He and Qian, Wei},
  booktitle={2019 41st annual international conference of the IEEE engineering in medicine and biology society (EMBC)},
  pages={2045--2048},
  year={2019},
  organization={IEEE}
}

@article{ref113,
  title={Kolmogorov-Arnold Vision Transformer for Image Reconstruction in Lung Electrical Impedance Tomography},
  author={Amin, Ibrar and Shi, Shuaikai and AlMarzouqi, Hasan and Aung, Zeyar and Ahmed, Waqar and Liatsis, Panos},
  journal={IEEE Open Journal of the Computer Society},
  year={2025},
  publisher={IEEE}
}

@inproceedings{ref114,
  title={Diabetic retinopathy detection using CNN, transformer and MLP based architectures},
  author={Kumar, Nikhil Sathya and Karthikeyan, B Ramaswamy},
  booktitle={2021 International Symposium on Intelligent Signal Processing and Communication Systems (ISPACS)},
  pages={1--2},
  year={2021},
  organization={IEEE}
}

@data{ref115,
doi = {10.21227/s3g7-st65},
url = {https://dx.doi.org/10.21227/s3g7-st65},
author = {Samiksha Pachade and Prasanna Porwal and Dhanshree Thulkar and Manesh Kokare and Girish Deshmukh and Vivek Sahasrabuddhe and Luca Giancardo and Gwenolé Quellec and Fabrice Mériaudeau},
publisher = {IEEE Dataport},
title = {Retinal Fundus Multi-disease Image Dataset (RFMiD)},
year = {2020}
}

@article{ref116,
  title={Detecting severity of diabetic retinopathy from fundus images using ensembled transformers},
  author={Adak, Chandranath and Karkera, Tejas and Chattopadhyay, Soumi and Saqib, Muhammad},
  journal={arXiv preprint arXiv:2301.00973},
  year={2023}
}

@article{ref117,
  title={Classification of diabetic retinopathy severity in fundus images using the vision transformer and residual attention},
  author={Gu, Zongyun and Li, Yan and Wang, Zijian and Kan, Junling and Shu, Jianhua and Wang, Qing},
  journal={Computational Intelligence and Neuroscience},
  volume={2023},
  number={1},
  pages={1305583},
  year={2023},
  publisher={Wiley Online Library}
}

@article{ref118,
  title={Vision Transformers in medical computer vision—A contemplative retrospection},
  author={Parvaiz, Arshi and Khalid, Muhammad Anwaar and Zafar, Rukhsana and Ameer, Huma and Ali, Muhammad and Fraz, Muhammad Moazam},
  journal={Engineering Applications of Artificial Intelligence},
  volume={122},
  pages={106126},
  year={2023},
  publisher={Elsevier}
}

@article{ref119,
  title={Applications of vision transformers in retinal imaging: A systematic review},
  author={Ye, En Zhou and Ye, Joseph and Ye, En Hui},
  journal={Authorea Preprints},
  year={2023},
  publisher={Authorea}
}

@article{ref120,
  title={Retinal vascular manifestations of metabolic disorders},
  author={Nguyen, Thanh T and Wong, Tien Y},
  journal={Trends in endocrinology \& metabolism},
  volume={17},
  number={7},
  pages={262--268},
  year={2006},
  publisher={Elsevier}
}

@article{ref121,
  title={Comprehensive assessment of systemic arteriosclerosis in relation to the ocular resistive index in acute coronary syndrome patients},
  author={Ebuchi, Yasunari and Nagaoka, Taiji and Fukamachi, Daisuke and Kojima, Keisuke and Akutsu, Naotaka and Murata, Nobuhiro and Saito, Yuki and Kitano, Daisuke and Yokota, Harumasa and Yamagami, Satoru and others},
  journal={Scientific Reports},
  volume={12},
  number={1},
  pages={2321},
  year={2022},
  publisher={Nature Publishing Group UK London}
}

@article{ref122,
  title={Review and comparison of retinal vessel calibre and geometry software and their application to diabetes, cardiovascular disease, and dementia},
  author={Brazionis, Laima and Quinn, Nicola and Dabbah, Sami and Ryan, Chris D and M{\o}ller, Dennis M and Richardson, Hilary and Keech, Anthony C and Januszewski, Andrzej S and Grauslund, Jakob and Rasmussen, Malin Lundberg and others},
  journal={Graefe's Archive for Clinical and Experimental Ophthalmology},
  volume={261},
  number={8},
  pages={2117--2133},
  year={2023},
  publisher={Springer}
}

@article{ref122a,
  title={Fractal analysis of region-based vascular change in the normal and non-proliferative diabetic retina},
  author={Avakian, Arpenik and Kalina, Robert E and Helene Sage, E and Rambhia, Avni H and Elliott, Katherine E and Chuang, Elaine L and Clark, John I and Hwang, Jenq-Neng and Parsons-Wingerter, Patricia},
  journal={Current eye research},
  volume={24},
  number={4},
  pages={274--280},
  year={2002},
  publisher={Taylor \& Francis}
}

@article{ref122b,
  title={A novel method for the automatic grading of retinal vessel tortuosity},
  author={Grisan, Enrico and Foracchia, Marco and Ruggeri, Alfredo},
  journal={IEEE transactions on medical imaging},
  volume={27},
  number={3},
  pages={310--319},
  year={2008},
  publisher={IEEE}
}

@misc{ref122c,
  title={Commentary on Lavia et al: progress of optical coherence tomography angiography for visualizing human retinal vasculature},
  author={Curcio, Christine A and Kar, Deepayan},
  journal={Retina},
  volume={39},
  number={2},
  pages={223--225},
  year={2019},
  publisher={LWW}
}

@article{ref123,
  title={Retinal arteriolar narrowing and left ventricular remodeling: the multi-ethnic study of atherosclerosis},
  author={Cheung, Ning and Bluemke, David A and Klein, Ronald and Sharrett, A Richey and Islam, FM Amirul and Cotch, Mary Frances and Klein, Barbara EK and Criqui, Michael H and Wong, Tien Yin},
  journal={Journal of the American College of Cardiology},
  volume={50},
  number={1},
  pages={48--55},
  year={2007},
  publisher={American College of Cardiology Foundation Washington, DC}
}

@article{ref124,
  title={Retinal vessel diameter as a clinical predictor of diabetic retinopathy progression: time to take out the measuring tape},
  author={Wong, Tien Yin},
  journal={Archives of ophthalmology},
  volume={129},
  number={1},
  pages={95--96},
  year={2011},
  publisher={American Medical Association}
}

@article{ref125,
  title={Retinal vascular tortuosity, blood pressure, and cardiovascular risk factors},
  author={Cheung, Carol Yim-lui and Zheng, Yingfeng and Hsu, Wynne and Lee, Mong Li and Lau, Qiangfeng Peter and Mitchell, Paul and Wang, Jie Jin and Klein, Ronald and Wong, Tien Yin},
  journal={Ophthalmology},
  volume={118},
  number={5},
  pages={812--818},
  year={2011},
  publisher={Elsevier}
}

@article{ref126,
  title={Retinal arteriolar tortuosity and cardiovascular risk factors in a multi-ethnic population study of 10-year-old children; the Child Heart and Health Study in England (CHASE)},
  author={Owen, Christopher G and Rudnicka, Alicja R and Nightingale, Claire M and Mullen, Robert and Barman, Sarah A and Sattar, Naveed and Cook, Derek G and Whincup, Peter H},
  journal={Arteriosclerosis, thrombosis, and vascular biology},
  volume={31},
  number={8},
  pages={1933--1938},
  year={2011},
  publisher={Am Heart Assoc}
}

@article{ref127,
  title={Automated characterization of blood vessels as arteries and veins in retinal images},
  author={Mirsharif, Qazaleh and Tajeripour, Farshad and Pourreza, Hamidreza},
  journal={Computerized Medical Imaging and Graphics},
  volume={37},
  number={7-8},
  pages={607--617},
  year={2013},
  publisher={Elsevier}
}

@article{ref128,
  title={An automatic graph-based approach for artery/vein classification in retinal images},
  author={Dashtbozorg, Behdad and Mendon{\c{c}}a, Ana Maria and Campilho, Aur{\'e}lio},
  journal={IEEE Transactions on Image Processing},
  volume={23},
  number={3},
  pages={1073--1083},
  year={2013},
  publisher={IEEE}
}

@article{ref129,
  title={Retinal artery-vein classification via topology estimation},
  author={Estrada, Rolando and Allingham, Michael J and Mettu, Priyatham S and Cousins, Scott W and Tomasi, Carlo and Farsiu, Sina},
  journal={IEEE transactions on medical imaging},
  volume={34},
  number={12},
  pages={2518--2534},
  year={2015},
  publisher={IEEE}
}

@inproceedings{ref130,
  title={U-net: Convolutional networks for biomedical image segmentation},
  author={Ronneberger, Olaf and Fischer, Philipp and Brox, Thomas},
  booktitle={Medical image computing and computer-assisted intervention--MICCAI 2015: 18th international conference, Munich, Germany, October 5-9, 2015, proceedings, part III 18},
  pages={234--241},
  year={2015},
  organization={Springer}
}

@article{ref131,
  title={Artery/vein classification using reflection features in retina fundus images},
  author={Huang, Fan and Dashtbozorg, Behdad and Romeny, Bart M ter Haar},
  journal={Machine Vision and Applications},
  volume={29},
  number={1},
  pages={23--34},
  year={2018},
  publisher={Springer}
}

@article{ref132,
  title={Automated method for retinal artery/vein separation via graph search metaheuristic approach},
  author={Srinidhi, Chetan L and Aparna, P and Rajan, Jeny},
  journal={IEEE Transactions on Image Processing},
  volume={28},
  number={6},
  pages={2705--2718},
  year={2019},
  publisher={IEEE}
}

@article{ref133,
  title={RAVIR: A dataset and methodology for the semantic segmentation and quantitative analysis of retinal arteries and veins in infrared reflectance imaging},
  author={Hatamizadeh, Ali and Hosseini, Hamid and Patel, Niraj and Choi, Jinseo and Pole, Cameron C and Hoeferlin, Cory M and Schwartz, Steven D and Terzopoulos, Demetri},
  journal={IEEE Journal of Biomedical and Health Informatics},
  volume={26},
  number={7},
  pages={3272--3283},
  year={2022},
  publisher={IEEE}
}

@article{ref133a,
  title={Phenotypic screening and genetic insights for predicting major vascular-related diseases using retinal imaging},
  author={Lu, Menglin and Mao, Yiheng and Zhu, Hui and Xu, Yesheng and Yao, Yu-Feng and Wu, Fei and Huang, Zhengxing},
  journal={NPJ Digital Medicine},
  volume={8},
  number={1},
  pages={437},
  year={2025},
  publisher={Nature Publishing Group UK London}
}

@article{ref133b,
  title={Deep phenotyping of health--disease continuum in the Human Phenotype Project},
  author={Reicher, Lee and Shilo, Smadar and Godneva, Anastasia and Lutsker, Guy and Zahavi, Liron and Shoer, Saar and Krongauz, David and Rein, Michal and Kohn, Sarah and Segev, Tomer and others},
  journal={Nature Medicine},
  pages={1--13},
  year={2025},
  publisher={Nature Publishing Group US New York}
}

@article{ref133c,
  title={Phenome-wide associations of human aging uncover sex-specific dynamics},
  author={Reicher, Lee and Bar, Noam and Godneva, Anastasia and Reisner, Yotam and Zahavi, Liron and Shahaf, Nir and Dhir, Raja and Weinberger, Adina and Segal, Eran},
  journal={Nature Aging},
  volume={4},
  number={11},
  pages={1643--1655},
  year={2024},
  publisher={Nature Publishing Group US New York}
}

@article{ref133d,
  title={Plasma proteomics links brain and immune system aging with healthspan and longevity},
  author={Oh, Hamilton Se-Hwee and Le Guen, Yann and Rappoport, Nimrod and Urey, Deniz Yagmur and Farinas, Amelia and Rutledge, Jarod and Channappa, Divya and Wagner, Anthony D and Mormino, Elizabeth and Brunet, Anne and others},
  journal={Nature Medicine},
  pages={1--9},
  year={2025},
  publisher={Nature Publishing Group US New York}
}

@article{ref133e,
  title={Next-generation phenotyping of inherited retinal diseases from multimodal imaging with Eye2Gene},
  author={Pontikos, Nikolas and Woof, William A and Lin, Siying and Ghoshal, Biraja and Mendes, Bernardo S and Veturi, Advaith and Nguyen, Quang and Javanmardi, Behnam and Georgiou, Michalis and Hustinx, Alexander and others},
  journal={Nature Machine Intelligence},
  pages={1--12},
  year={2025},
  publisher={Nature Publishing Group UK London}
}

@article{ref133f,
  title={A deep learning system for detecting systemic lupus erythematosus from retinal images},
  author={Li, Tingyao and Lin, Shiqun and Guan, Zhouyu and Zhou, Yukun and Zeng, Dian and Wang, Zheyuan and Zhou, Yan and Fang, Pinqi and Yu, Shujie and Liu, Ruhan and others},
  journal={Cell Reports Medicine},
  year={2025},
  publisher={Elsevier}
}

@misc{ref134,
  author       = {D. A. S. Ph.D},
  title        = {The Eyes Are The Windows To The Soul},
  year         = {2024},
  howpublished = {\url{https://www.mentalhealth.com/library/the-eyes-are-the-windows-to-the-soul}},
  note         = {Accessed: 27 April 2025},
  publisher    = {MentalHealth.com}
}

@misc{ref135,
  author       = {{XLab Health}},
  title        = {The Soul and Eyes: An Influential Psychiatrist Analyses Whether the Eyes Are the Window to the Soul and Why It Is Important to Take Care of Eye Health},
  year         = {2024},
  howpublished = {\url{https://xlab.health/the-soul-and-eyes-an-influential-psychiatrist-analyses-whether-the-eyes-are-the-window-to-the-soul-and-why-it-is-important-to-take-care-of-eye-health/}},
  note         = {Accessed: 27 April 2025},
  publisher    = {XLab Health}
}

@article{ref136,
  title={2013 ACC/AHA guideline on the assessment of cardiovascular risk: a report of the American College of Cardiology/American Heart Association Task Force on Practice Guidelines},
  author={Goff Jr, David C and Lloyd-Jones, Donald M and Bennett, Glen and Coady, Sean and D’agostino, Ralph B and Gibbons, Raymond and Greenland, Philip and Lackland, Daniel T and Levy, Daniel and O’donnell, Christopher J and others},
  journal={Circulation},
  volume={129},
  number={25\_suppl\_2},
  pages={S49--S73},
  year={2014},
  publisher={Lippincott Williams \& Wilkins Hagerstown, MD}
}

@article{ref137,
  title={Frequency and practice-level variation in inappropriate aspirin use for the primary prevention of cardiovascular disease: insights from the National Cardiovascular Disease Registry’s Practice Innovation and Clinical Excellence registry},
  author={Hira, Ravi S and Kennedy, Kevin and Nambi, Vijay and Jneid, Hani and Alam, Mahboob and Basra, Sukhdeep S and Ho, P Michael and Deswal, Anita and Ballantyne, Christie M and Petersen, Laura A and others},
  journal={Journal of the American College of Cardiology},
  volume={65},
  number={2},
  pages={111--121},
  year={2015},
  publisher={American College of Cardiology Foundation Washington, DC}
}

@article{ref138,
  title={Retinal microvasculature as a model to study the manifestations of hypertension},
  author={Cheung, Carol Yim-lui and Ikram, M Kamran and Sabanayagam, Charumathi and Wong, Tien Yin},
  journal={Hypertension},
  volume={60},
  number={5},
  pages={1094--1103},
  year={2012},
  publisher={Lippincott Williams \& Wilkins Hagerstown, MD}
}

@article{ref138a,
  title={Prediction of incident stroke events based on retinal vessel caliber: a systematic review and individual-participant meta-analysis},
  author={McGeechan, Kevin and Liew, Gerald and Macaskill, Petra and Irwig, Les and Klein, Ronald and Klein, Barbara EK and Wang, Jie Jin and Mitchell, Paul and Vingerling, Johannes R and De Jong, Paulus TVM and others},
  journal={American journal of epidemiology},
  volume={170},
  number={11},
  pages={1323--1332},
  year={2009},
  publisher={Oxford University Press}
}

@article{ref139,
  title={Hypertensive retinopathy can predict stroke: A systematic review and meta-analysis based on observational studies: Running title: HTR and Stroke: Meta-analysis},
  author={Wang, Zhe and Feng, Liyuan and Wu, Mei and Ding, Fengxing and Liu, Chen and Xie, Guangmei and Ma, Bin},
  journal={Journal of Stroke and Cerebrovascular Diseases},
  pages={107953},
  year={2024},
  publisher={Elsevier}
}

@article{ref139a,
  title={Artificial intelligence-enabled retinal vasculometry for prediction of circulatory mortality, myocardial infarction and stroke},
  author={Rudnicka, Alicja Regina and Welikala, Roshan and Barman, Sarah and Foster, Paul J and Luben, Robert and Hayat, Shabina and Khaw, Kay-Tee and Whincup, Peter and Strachan, David and Owen, Christopher G},
  journal={British Journal of Ophthalmology},
  volume={106},
  number={12},
  pages={1722--1729},
  year={2022},
  publisher={BMJ Publishing Group Ltd}
}

@article{ref139b,
  title={Prediction of major adverse cardiovascular events from retinal, clinical, and genomic data in individuals with type 2 diabetes: a population cohort study},
  author={Mordi, Ify R and Trucco, Emanuele and Syed, Mohammad Ghouse and MacGillivray, Tom and Nar, Adi and Huang, Yu and George, Gittu and Hogg, Stephen and Radha, Venkatesan and Prathiba, Vijayaraghavan and others},
  journal={Diabetes Care},
  volume={45},
  number={3},
  pages={710--716},
  year={2022},
  publisher={American Diabetes Association}
}

@article{ref139c,
  title={Retinal photograph-based deep learning predicts biological age, and stratifies morbidity and mortality risk},
  author={Nusinovici, Simon and Rim, Tyler Hyungtaek and Yu, Marco and Lee, Geunyoung and Tham, Yih-Chung and Cheung, Ning and Chong, Crystal Chun Yuen and Da Soh, Zhi and Thakur, Sahil and Lee, Chan Joo and others},
  journal={Age and ageing},
  volume={51},
  number={4},
  pages={afac065},
  year={2022},
  publisher={Oxford University Press}
}

@article{ref139d,
  title={Retinal age gap as a predictive biomarker for mortality risk},
  author={Zhu, Zhuoting and Shi, Danli and Guankai, Peng and Tan, Zachary and Shang, Xianwen and Hu, Wenyi and Liao, Huan and Zhang, Xueli and Huang, Yu and Yu, Honghua and others},
  journal={British Journal of Ophthalmology},
  volume={107},
  number={4},
  pages={547--554},
  year={2023},
  publisher={BMJ Publishing Group Ltd}
}

@article{ref139e,
  title={Cardiovascular disease diagnosis from DXA scan and retinal images using deep learning},
  author={Al-Absi, Hamada RH and Islam, Mohammad Tariqul and Refaee, Mahmoud Ahmed and Chowdhury, Muhammad EH and Alam, Tanvir},
  journal={Sensors},
  volume={22},
  number={12},
  pages={4310},
  year={2022},
  publisher={MDPI}
}

@article{ref139f,
  title={Can deep learning on retinal images augment known risk factors for cardiovascular disease prediction in diabetes? A prospective cohort study from the national screening programme in Scotland},
  author={Mellor, Joseph and Jiang, Wenhua and Fleming, Alan and McGurnaghan, Stuart J and Blackbourn, Luke and Styles, Caroline and Storkey, Amos J and McKeigue, Paul M and Colhoun, Helen M and Scottish Diabetes Research Network Epidemiology Group and others},
  journal={International journal of medical informatics},
  volume={175},
  pages={105072},
  year={2023},
  publisher={Elsevier}
}

@article{ref139g,
  title={Development and validation of retinal vasculature nomogram in suspected angina due to coronary artery disease},
  author={Zhong, Pingting and Qin, Jie and Li, Zhixi and Jiang, Lei and Peng, Qingsheng and Huang, Manqing and Lin, Yingwen and Liu, Baoyi and Li, Cong and Wu, Qiaowei and others},
  journal={Journal of atherosclerosis and thrombosis},
  volume={29},
  number={5},
  pages={579--596},
  year={2022},
  publisher={Japan Atherosclerosis Society}
}

@article{ref142,
  title={IDF Diabetes Atlas: Global, regional and country-level diabetes prevalence estimates for 2021 and projections for 2045},
  author={Sun, Hong and Saeedi, Pouya and Karuranga, Suvi and Pinkepank, Moritz and Ogurtsova, Katherine and Duncan, Bruce B and Stein, Caroline and Basit, Abdul and Chan, Juliana CN and Mbanya, Jean Claude and others},
  journal={Diabetes research and clinical practice},
  volume={183},
  pages={109119},
  year={2022},
  publisher={Elsevier}
}

@article{ref143,
  title={Complication of diabetes mellitus: Microvascular and macrovascular complications},
  author={Bereda, G},
  journal={Int. J. Diabetes},
  volume={3},
  pages={123--128},
  year={2022}
}

@article{ref144,
  title={Retinal OCT-Derived Texture Features as Potential Biomarkers for Early Diagnosis and Progression of Diabetic Retinopathy},
  author={Oliveira, Sara and Guimar{\~a}es, Pedro and Campos, Elisa Juli{\~a}o and Fernandes, Rosa and Martins, Jo{\~a}o and Castelo-Branco, Miguel and Serranho, Pedro and Matafome, Paulo and Bernardes, Rui and Ambr{\'o}sio, Ant{\'o}nio Francisco},
  journal={Investigative Ophthalmology \& Visual Science},
  volume={66},
  number={1},
  pages={7--7},
  year={2025},
  publisher={The Association for Research in Vision and Ophthalmology}
}

@article{ref145,
  title={Retinal microvascular calibre and risk of diabetes mellitus: a systematic review and participant-level meta-analysis},
  author={Sabanayagam, Charumathi and Lye, Weng Kit and Klein, Ronald and Klein, Barbara EK and Cotch, Mary Frances and Wang, Jie Jin and Mitchell, Paul and Shaw, Jonathan E and Selvin, Elizabeth and Sharrett, A Richey and others},
  journal={Diabetologia},
  volume={58},
  pages={2476--2485},
  year={2015},
  publisher={Springer}
}

@article{ref146,
  title={Circulating biomarkers of diabetic retinopathy: an overview based on physiopathology},
  author={Sim{\'o}-Servat, Olga and Sim{\'o}, Rafael and Hern{\'a}ndez, Cristina},
  journal={Journal of Diabetes Research},
  volume={2016},
  number={1},
  pages={5263798},
  year={2016},
  publisher={Wiley Online Library}
}

@article{ref147,
  title={Frontiers in Understanding the Pathological Mechanism of Diabetic Retinopathy},
  author={Zhan, Lei},
  journal={Medical Science Monitor: International Medical Journal of Experimental and Clinical Research},
  volume={29},
  pages={e939658--1},
  year={2023},
  publisher={International Scientific Information, Inc.}
}

@article{ref148,
  title={Retinopathy progression and the risk of end-stage kidney disease: results from a longitudinal Japanese cohort of 232 patients with type 2 diabetes and biopsy-proven diabetic kidney disease},
  author={Yamanouchi, Masayuki and Mori, Mikiro and Hoshino, Junichi and Kinowaki, Keiichi and Fujii, Takeshi and Ohashi, Kenichi and Furuichi, Kengo and Wada, Takashi and Ubara, Yoshifumi},
  journal={BMJ open diabetes research \& care},
  volume={7},
  number={1},
  year={2019},
  publisher={American Diabetes Association}
}

@article{ref149,
  title={Correlation between diabetic retinopathy and diabetic nephropathy: a two-sample Mendelian randomization study},
  author={Fang, Jiaxi and Luo, Chuxuan and Zhang, Di and He, Qiang and Liu, Lin},
  journal={Frontiers in Endocrinology},
  volume={14},
  pages={1265711},
  year={2023},
  publisher={Frontiers Media SA}
}

@article{ref150,
  title={Macular thickness decrease in asymptomatic subjects at high genetic risk of developing Alzheimer’s disease: an OCT study},
  author={L{\'o}pez-Cuenca, In{\'e}s and de Hoz, Rosa and Salobrar-Garc{\'\i}a, Elena and Elvira-Hurtado, Lorena and Rojas, Pilar and Fern{\'a}ndez-Albarral, Jos{\'e} A and Barabash, Ana and Salazar, Juan J and Ram{\'\i}rez, Ana I and Ram{\'\i}rez, Jos{\'e} M},
  journal={Journal of Clinical Medicine},
  volume={9},
  number={6},
  pages={1728},
  year={2020},
  publisher={MDPI}
}

@article{ref151,
  title={Retinal ganglion cell complex in Alzheimer Disease: comparing Ganglion Cell Complex and Central Macular Thickness in Alzheimer Disease and healthy subjects using spectral domain-optical coherence tomography},
  author={Farzinvash, Zahra and Abutorabi-Zarchi, Marzie and Manaviat, Masoudreza and Mehrjerdi, Habib Zare},
  journal={Basic and clinical neuroscience},
  volume={13},
  number={5},
  pages={675},
  year={2022}
}

@article{ref152,
  title={Retinal vascular fractals and cognitive impairment},
  author={Ong, Yi-Ting and Hilal, Saima and Cheung, Carol Yim-lui and Xu, Xin and Chen, Christopher and Venketasubramanian, Narayanaswamy and Wong, Tien Yin and Ikram, Mohammad Kamran},
  journal={Dementia and geriatric cognitive disorders extra},
  volume={4},
  number={2},
  pages={305--313},
  year={2014},
  publisher={S. Karger AG Basel, Switzerland}
}

@article{ref153,
  title={Microvascular network alterations in the retina of patients with Alzheimer's disease},
  author={Cheung, Carol Yim-lui and Ong, Yi Ting and Ikram, M Kamran and Ong, Shin Yeu and Li, Xiang and Hilal, Saima and Catindig, Joseree-Ann S and Venketasubramanian, Narayanaswamy and Yap, Philip and Seow, Dennis and others},
  journal={Alzheimer's \& Dementia},
  volume={10},
  number={2},
  pages={135--142},
  year={2014},
  publisher={Elsevier}
}

@article{ref154,
  title={Retinal biomarkers for Alzheimer’s disease and vascular cognitive impairment and dementia (VCID): implication for early diagnosis and prognosis},
  author={Czak{\'o}, Cecilia and Kov{\'a}cs, Tibor and Ungvari, Zoltan and Csiszar, Anna and Yabluchanskiy, Andriy and Conley, Shannon and Csipo, Tamas and Lipecz, Agnes and Horv{\'a}th, Hajnalka and S{\'a}ndor, G{\'a}bor L{\'a}szl{\'o} and others},
  journal={Geroscience},
  volume={42},
  pages={1499--1525},
  year={2020},
  publisher={Springer}
}

@article{ref154a,
  title={A deep learning model for detection of Alzheimer's disease based on retinal photographs: a retrospective, multicentre case-control study},
  author={Cheung, Carol Y and Ran, An Ran and Wang, Shujun and Chan, Victor TT and Sham, Kaiser and Hilal, Saima and Venketasubramanian, Narayanaswamy and Cheng, Ching-Yu and Sabanayagam, Charumathi and Tham, Yih Chung and others},
  journal={The Lancet Digital Health},
  volume={4},
  number={11},
  pages={e806--e815},
  year={2022},
  publisher={Elsevier}
}

@article{ref155,
  title={Retinal thinning and correlation with functional disability in patients with Parkinson's disease},
  author={Satue, M and Seral, M and Otin, S and Alarcia, R and Herrero, R and Bambo, MP and Fuertes, MI and Pablo, LE and Garcia-Martin, EJBJOO},
  journal={British Journal of Ophthalmology},
  volume={98},
  number={3},
  pages={350--355},
  year={2014},
  publisher={BMJ Publishing Group Ltd}
}

@article{ref156,
  title={Retinal functional and structural changes in patients with Parkinson’s disease},
  author={Elanwar, Rehab and Al Masry, Hatem and Ibrahim, Amna and Hussein, Mona and Ibrahim, Sahar and Masoud, Mohammed M},
  journal={BMC neurology},
  volume={23},
  number={1},
  pages={330},
  year={2023},
  publisher={Springer}
}

@article{ref156a,
  title={EM-COGLOAD: An investigation into age and cognitive load detection using eye tracking and deep learning},
  author={Miles, Gabriella and Smith, Melvyn and Zook, Nancy and Zhang, Wenhao},
  journal={Computational and Structural Biotechnology Journal},
  volume={24},
  pages={264--280},
  year={2024},
  publisher={Elsevier}
}

@article{ref156b,
  title={Artificial intelligence in eye movements analysis for Alzheimer’s disease early diagnosis},
  author={Maleki, Shadi Farabi and Yousefi, Milad and Sobhi, Navid and Jafarizadeh, Ali and Alizadehsani, Roohallah and Gorriz-Saez, Juan Manuel},
  journal={Current Alzheimer Research},
  volume={21},
  number={3},
  pages={155--165},
  year={2024},
  publisher={Bentham Science Publishers}
}

@article{ref156c,
  title={Deep learning-based eye-tracking analysis for diagnosis of Alzheimer's disease using 3D comprehensive visual stimuli},
  author={Zuo, Fangyu and Jing, Peiguang and Sun, Jinglin and Duan, Jizhong and Ji, Yong and Liu, Yu},
  journal={IEEE Journal of Biomedical and Health Informatics},
  volume={28},
  number={5},
  pages={2781--2793},
  year={2024},
  publisher={IEEE}
}

@article{ref157,
  title={Retinal layer thinning for monitoring disease-modifying treatment in relapsing multiple sclerosis—Evidence for applying a rebaselining concept},
  author={Bsteh, Gabriel and Hegen, Harald and Krajnc, Nik and F{\"o}ttinger, Fabian and Altmann, Patrick and Auer, Michael and Berek, Klaus and Kornek, Barbara and Leutmezer, Fritz and Macher, Stefan and others},
  journal={Multiple Sclerosis Journal},
  volume={30},
  number={9},
  pages={1128--1138},
  year={2024},
  publisher={SAGE Publications Sage UK: London, England}
}

@article{ref157a,
  title={Machine learning in diagnosis and disability prediction of multiple sclerosis using optical coherence tomography},
  author={Montol{\'\i}o, Alberto and Mart{\'\i}n-Gallego, Alejandro and Cego{\~n}ino, Jos{\'e} and Orduna, Elvira and Vilades, Elisa and Garcia-Martin, Elena and Del Palomar, Amaya P{\'e}rez},
  journal={Computers in Biology and Medicine},
  volume={133},
  pages={104416},
  year={2021},
  publisher={Elsevie}
}

@article{ref157b,
  title={Early diagnosis of multiple sclerosis using swept-source optical coherence tomography and convolutional neural networks trained with data augmentation},
  author={Lopez-Dorado, Almudena and Ortiz, Miguel and Satue, Maria and Rodrigo, Maria J and Barea, Rafael and Sanchez-Morla, Eva M and Cavaliere, Carlo and Rodriguez-Ascariz, Jose M and Orduna-Hospital, Elvira and Boquete, Luciano and others},
  journal={Sensors},
  volume={22},
  number={1},
  pages={167},
  year={2021},
  publisher={Mdpi}
}

@article{ref157c,
  title={Retinal vascular calibers and the risk of intracerebral hemorrhage and cerebral infarction: the Rotterdam Study},
  author={Wieberdink, Renske G and Ikram, M Kamran and Koudstaal, Peter J and Hofman, Albert and Vingerling, Johannes R and Breteler, Monique MB},
  journal={Stroke},
  volume={41},
  number={12},
  pages={2757--2761},
  year={2010},
  publisher={Lippincott Williams \& Wilkins}
}

@article{ref157d,
  title={Retinal microvascular signs and risk of stroke: the Multi-Ethnic Study of Atherosclerosis (MESA)},
  author={Kawasaki, Ryo and Xie, Jing and Cheung, Ning and Lamoureux, Ecosse and Klein, Ronald and Klein, Barbara EK and Cotch, Mary Frances and Sharrett, A Richey and Shea, Steven and Wong, Tien Y and others},
  journal={Stroke},
  volume={43},
  number={12},
  pages={3245--3251},
  year={2012},
  publisher={Lippincott Williams \& Wilkins Hagerstown, MD}
}

@article{ref157e,
  title={The association between retinal vasculature changes and stroke: a literature review and meta-analysis},
  author={Wu, Hui-Qun and Wu, Huan and Shi, Li-Li and Yu, Li-Yuan and Wang, Li-Yuan and Chen, Ya-Lan and Geng, Jin-Song and Shi, Jian and Jiang, Kui and Dong, Jian-Cheng},
  journal={International journal of ophthalmology},
  volume={10},
  number={1},
  pages={109},
  year={2017}
}

@article{ref157f,
  title={Retinal vascular fingerprints predict incident stroke: findings from the UK Biobank cohort study},
  author={Yusufu, Mayinuer and Friedman, David S and Kang, Mengtian and Padhye, Ambhruni and Shang, Xianwen and Zhang, Lei and Shi, Danli and He, Mingguang},
  journal={Heart},
  volume={111},
  number={7},
  pages={306--313},
  year={2025},
  publisher={BMJ Publishing Group Ltd and British Cardiovascular Society}
}

@article{ref157g,
  title={A deep learning system for detecting silent brain infarction and predicting stroke risk},
  author={Jiang, Nan and Ji, Hongwei and Guan, Zhouyu and Pan, Yuesong and Deng, Chenxin and Guo, Yuchen and Liu, Dan and Chen, Tingli and Wang, Shiyu and Wu, Yilan and others},
  journal={Nature Biomedical Engineering},
  pages={1--13},
  year={2025},
  publisher={Nature Publishing Group UK London}
}

@article{ref159,
  title={Health system-scale language models are all-purpose prediction engines},
  author={Jiang, Lavender Yao and Liu, Xujin Chris and Nejatian, Nima Pour and Nasir-Moin, Mustafa and Wang, Duo and Abidin, Anas and Eaton, Kevin and Riina, Howard Antony and Laufer, Ilya and Punjabi, Paawan and others},
  journal={Nature},
  volume={619},
  number={7969},
  pages={357--362},
  year={2023},
  publisher={Nature Publishing Group UK London}
}

@article{ref160,
  title={Llava-med: Training a large language-and-vision assistant for biomedicine in one day},
  author={Li, Chunyuan and Wong, Cliff and Zhang, Sheng and Usuyama, Naoto and Liu, Haotian and Yang, Jianwei and Naumann, Tristan and Poon, Hoifung and Gao, Jianfeng},
  journal={Advances in Neural Information Processing Systems},
  volume={36},
  year={2024}
}

@article{ref161,
  title={A foundational multimodal vision language AI assistant for human pathology},
  author={Lu, Ming Y and Chen, Bowen and Williamson, Drew FK and Chen, Richard J and Ikamura, Kenji and Gerber, Georg and Liang, Ivy and Le, Long Phi and Ding, Tong and Parwani, Anil V and others},
  journal={arXiv preprint arXiv:2312.07814},
  year={2023}
}
\end{document}